\documentclass[aps,prd,superscriptaddress,eqsecnum]{revtex4}
\usepackage{subfigure}
\usepackage{color}
\usepackage{amssymb}
\usepackage[dvipdfmx]{graphicx}
\begin{document}

\title{Hybrid chiral condensate in the external magnetic field}
\author{Kazuya Nishiyama}
\affiliation{Department of Physics, Kyoto University, Kyoto 606-8502, Japan}
\author{Shintaro Karasawa}
\affiliation{Department of Physics, Kyoto University, Kyoto 606-8502, Japan}
\author{Toshitaka Tatsumi}
\affiliation{Department of Physics, Kyoto University, Kyoto 606-8502, Japan}

\begin{abstract}
We study the phase diagram of the Nambu-Jona-Lasinio model in the external magnetic field within the mean-field approximation, taking into  account the inhomogeneous chiral condensate. 
It is shown that there appears a new type of the chiral condensate, endowed with two features of real kink crystal and dual chiral density wave, in the magnetic field.
We also show that there are first order phase transitions between different inhomogeneous phases in the presence of magnetic field.

\end{abstract}
                                          
\maketitle

\section{Introduction\label{sec:intro}}

In the last decade, the possible appearance of the inhomogeneous chiral phase in the QCD phase diagram has been studied
\cite{Nakano:2004cd,Tatsumi:2014wka,Schon:2000qy,Thies:2003kk,Nickel:2009wj,Nickel:2009ke,Basar:2008ki,Basar:2009fg,Kojo:2009ha,Kojo:2011cn,Muller:2013tya,Buballa:2014tba,Carignano:2010ac,Carignano:2014jla,Carignano:2012sx,Karasawa:2013zsa,Maedan:2009yi,Abuki:2013vwa,Abuki:2013pla,Abuki:2011pf,Frolov:2010wn,Gubina:2012wp,Ebert:2014woa},
where the  quark condensate is spatially modulated.
For the analysis of the  inhomogeneous chiral phase, the chiral order parameter, 
$M({\bf x})=-2G\left[ \left< \overline \psi \psi\right> +i\left< \overline \psi i\gamma^5\tau^3\psi\right>\right]$,
has been used. 
Using the effective models of QCD including the Nambu-Jona-Lasinio  model or the Schwinger-Dyson approach,
there appears the inhomogeneous chiral phase in the vicinity of the chiral transition and its critical point is changed to be the Lifshitz point \cite{Nickel:2009ke}.
The dual chiral density wave (DCDW) or the real kink crystal (RKC) has been  often used as a typical condensate with one dimensional spatial modulation.
DCDW is a plane wave configuration, $ M({\bf x}) =me^{iqz}$, 
while RKC is a multi-soliton configuration,  $ M({\bf x}) =\frac{2m\nu}{1+\sqrt{\nu}}{\rm sn}\left( \frac{2mz}{1+\sqrt{\nu}}, \nu\right) $,
without the phase degree of freedom. 
In Ref. \cite{Nickel:2009wj}, it has been shown that one or two dimensional modulation can be embedded in 1+3 dimensions  by using the Lorentz boost.
The general solutions have been obtained by using the NJL$_2$ model in 1+1 dimensions \cite{Basar:2008ki}, which is called complex kink crystal. 
Higher dimensional modulations has been also considered in some studies \cite{Basar:2008ki,Basar:2009fg,Carignano:2012sx,Abuki:2011pf,Kojo:2011cn}.

In Refs. \cite{Nakano:2004cd,Nickel:2009wj,Nickel:2009ke,Muller:2013tya},
it has been shown that the inhomogeneous chiral phase can appear at low temperature and moderate density region as an intermediate phase prior to the chiral transition.
Such inhomogeneous phase has been also studied in condensed matter physics, e.g. spin or charge density wave and the FFLO state of the superconductivity \cite{Overhauser:1962zz,Pierls:1955,Fulde:1964zz,larkin:1964zz}.
The appearance of the inhomogeneous chiral phase in the QCD diagram has been extensively studied by using the various approaches, 
but there are few works about the external field, isospin asymmetry, and current quark mass.
These effects should be very important in realistic situations, especially for compact stars.
The effect of the current quark mass has been studied in some papers \cite{Nickel:2009wj,Maedan:2009yi,Karasawa:2013zsa} and change of the phase diagram has been figured out.

In this paper we consider the inhomogeneous chiral phase in the presence of the external magnetic field to figure out some magnetic properties.
The effect of the magnetic field is theoretically and phenomenologically interesting and important,
since quark matter is put into the strong magnetic field in compact stars or in heavy-ion collision process \cite{Skokov:2009qp}. 
QCD in the external magnetic field has recently attracted great attention,
and it has been shown that the magnetic field gives rise to various phenomena such as chiral magnetic effect \cite{Fukushima:2008xe},
magnetic catalysis \cite{Gusynin:1995nb,Suganuma:1990nn,Klevansky:1989vi}, magnetic inhibition \cite{Bali:2011qj,Fukushima:2012kc,Braun:2014fua,Mueller:2015fka}.
Lattice QCD simulations have been also performed to study the properties of the QCD vacuum in response to the magnetic field \cite{Bali:2011qj}; 
the effect of the magnetic field on the chiral transition or deconfinement has been studied at chemical potential $\mu=0$. 

The property of the inhomogeneous chiral condensate in the magnetic field has been first studied by Frolov et al. \cite{Frolov:2010wn}. 
They have found that the DCDW phase develops in a wide density region at $T=0$ under the magnetic field, 
and that some peculiar behaviours of the amplitude and of the wavevector can be seen due to the de Haas-van Alphen effect \cite{deHaas1936,Ebert:1999ht}.
However, they did not take into account the possibility of the RKC suggested to be favored in the absence of the magnetic field \cite{Nickel:2009wj}.

In this paper, we study the QCD phase diagram in the Nambu-Jona-Lasinio (NJL) model \cite{Nambu:1961tp,Hatsuda:1994pi,Klevansky:1992qe} in the magnetic field, taking into account both of the condensates.
We introduce the a new type of the condensate called hybrid chiral condensate (HCC) ,
$
 M({\bf x}) = \frac{2m\nu}{1+\sqrt{\nu}}{\rm sn}\left( \frac{2mz}{1+\sqrt{\nu}},\nu \right)e^{iqz}
$, which smoothly connects both DCDW and RKC by changing the modulus $\nu$ or the wavevector $q$, and demonstrate that the magnetic field favors the phase modulation: 
it is found through the analysis of the thermodynamic potential that the wavevector $q$ takes a nonzero value in the presence of the magnetic field,
and thus DCDW and RKC coexist as HCC in the weak magnetic field at moderate densities.
We shall see that the phase degree of freedom in HCC plays an important role in the presence of the magnetic field. 
The energy spectrum of the quark field becomes asymmetric in the presence of the magnetic field, which gives rise to anomalous quark number \cite{Tatsumi:2014wka}. 
Such spectral asymmetry is closely connected with chiral anomaly and moves the Lifshitz point to zero chemical potential $\mu=0$ .

We consider only the case of isospin symmetric matter ($\mu_u=\mu_d$) in the chiral limit($m_c=0$) for simplicity.
In  Sec. II we briefly summarize the general framework to deal with the inhomogeneous chiral phases in the presence of the magnetic field. 
We introduce HCC in Sec.III. 
Spectral asymmetry in the HCC phase and some topological features are also discussed there.
The phase diagram is presented in Sec.IV in the $B-\mu$ plane. Sec.V is devoted to concluding remarks.
The proper-time regularization method is given Appendix A. 
Some details about spectral asymmetry are presented in Appendix B and C. 
An expansion of the thermodynamic potential with respect to $B$ is given in Appendix D .

\section{Model and Energy spectrum\label{sec:model}}
Here we briefly summarize the general framework to get the quark energy spectrum in the inhomogeneous chiral phase in the presence of the magnetic field.
First, we consider the case of $N_f=N_c=1$ for simplicity.
The case of three colors and two flavors is also calculated in the same way. 
Taking the magnetic field $\bf B$ along , e.g., $z$ axis, the Lagrangian reads
\begin{eqnarray}
 L &=& \overline \psi\left[ -i\gamma^\mu D_\mu  +\frac{1+\gamma^5}{2}M +\frac{1-\gamma^5}{2}M^*\right] \psi -\frac{\left| M\right|^2}{4G},
\end{eqnarray}
within the mean-field approximation,  where $\psi$ is 4-dimensional spinor,
 $M$ is the order parameter of chiral transition, $M({\bf x})=-2G\left[ \left< \overline \psi \psi\right> +i\left< \overline \psi i\gamma^5\tau^3\psi\right>\right]$,
 ,$D_\mu$ is covariant derivative $D_\mu = \partial_\mu -ieA_\mu$.
We consider one dimensional modulation along the $z$ axis as well, $M({\bf x})=M(z)$.
It is assumed that magnetic field is uniform and parallel to modulation of the order parameter.
We shall see later that this orientation should be most favorable due to the topological aspects \cite{Tatsumi:2014wka}.
We combined Nickel's method \cite{Nickel:2009wj} and Frolov's method \cite{Frolov:2010wn} for obtaining the energy spectrum.

We choose the Landau gauge, $A^\mu=(0,{\bf A}),{\bf A}=(0,xB,0)$, and assume $eB>0$.
The Hamiltonian then renders 
\begin{eqnarray}
H_D &=& {\bf \alpha\cdot\Pi}+\gamma^{0}M\frac{1+\gamma^{5}}{2}+\gamma^{0}M^{*}\frac{1-\gamma^{5}}{2},
\end{eqnarray}
where ${\bf \Pi}$ is kinetic momentum, $\Pi_i = -i\partial_i + eA_i$.
The Hamiltonian satisfies the commutation relation, $[H_D,-i\partial_{y}]=[H_D,(\alpha_{\perp}\cdot\Pi_\perp)^{2}]=0$,
where ${\bf \alpha}_\perp=(\alpha_x,\alpha_y,0)$ and ${\bf \Pi}_\perp=(\Pi_x,\Pi_y,0)$ .
The eigenspinor of $-i\partial_{y}$ and $(\alpha_{\perp}\cdot\Pi_\perp)^{2}$ can be written as
 \begin{eqnarray}
\psi_{n,k}=\frac{1}{\sqrt{2\pi}}\left(eB \right)^{1/4}e^{iky}\left(\begin{array}{ccc}    c_{1}(z)u_{n-1}(\eta)\\
 ic_{2}(z)u_{n}(\eta) \\
   c_{3}(z)u_{n-1}(\eta)\\
 ic_{4}(z)u_{n}(\eta)\\
\end{array}
\right),
\end{eqnarray}
where $\eta=x\sqrt{eB}+k/\sqrt{eB}$,  $u_{n}(\eta)$ is the Hermite function \cite{Sokolov1986} which satisfies
$
 \left(\frac{\partial}{\partial\eta}+i\eta \right)u_n(\eta)=\sqrt{2n}u_{n-1}(\eta)
 $
 and
 $
 \left(\frac{\partial}{\partial\eta}-i\eta \right)u_{n-1}(\eta)=-\sqrt{2n}u_{n}(\eta),
$
and $n=0,1,2,....$ denotes the discrete Landau levels.
Using this eigenspinor, the Hartree-Fock equation $H_D\psi = {\cal E} \psi $ is reduced to
 \begin{eqnarray}
\left(\begin{array}{cccc}   -i\partial_z & 0 & M(z) & \sqrt{2eBn} \\
  0&i\partial_z &\sqrt{2eBn} & M^{*}(z)\\
  M^{*}(z)&\sqrt{2eBn} &i\partial_z  & 0 \\
\sqrt{2eBn}  & M(z)& 0 & -i\partial_z \\
\end{array}
\right)
\left(\begin{array}{c}    c_{1}(z)\\
 c_{2}(z) \\
   c_{3}(z)\\
 c_{4}(z)\\
\end{array}
\right)
=
{\cal E}
\left(\begin{array}{c}     c_{1}(z)\\
 c_{2}(z) \\
   c_{3}(z)\\
 c_{4}(z)\\
\end{array}
\right).
\label{eq:3dh}
\end{eqnarray}
for $n=1,2,...$, and 
 \begin{eqnarray}
\left(\begin{array}{cc}   
  i\partial_z & M^{*}(z)\\
 M(z) & -i\partial_z \\
\end{array}
\right)
\left(\begin{array}{c}
 c_{2}(z) \\
 c_{4}(z)\\
\end{array}
\right)
=
{\cal E}
\left(\begin{array}{c} 
 c_{2}(z) \\
 c_{4}(z)\\
\end{array}
\right).
\label{eq:1dh}
\end{eqnarray}
for $n=0$.
The latter equation (\ref{eq:1dh}) resembles the Bogoliubov-de Gennes (BdG) equation in 1+1 dimensions, 
while the former equation (\ref{eq:3dh}) is the same form as the one  without magnetic field. 
Thus the energy spectrum for these equations can be obtained once the corresponding one is given in the absence of the magnetic field.
The energy spectrum in the case of $B=0$ is simply written as ${\cal E}=\lambda_{\pm} \sqrt{1+p_{\perp}^2/{\lambda_\pm^2}}$ with 
the perpendicular component of the momentum, $p_{\perp}$ \cite{Nickel:2009wj} ,
where $\lambda_+$ is the eigenenergy  of the 1+1 dimensional Hartree-Fock equation \cite{Basar:2008ki}, 
  \begin{eqnarray}
\left(\begin{array}{cc}
i\partial_z & M(z)\\
M(z)^* & -i\partial_z
\end{array}
\right)\psi &=& \lambda_+\psi
,
 \end{eqnarray}
and $\lambda_-$ is defined as the eigenenergy for the complex conjugate transformation : $M({\bf x})\to M({\bf x})^*$.
Since Eq.~ (\ref{eq:3dh}) has a similar form to the usual Dirac equation with momentum $p_\perp$, the eigenvalue can be simply obtained by replacing $p_\perp$ by $\sqrt{2eBn}$.
 Thus we obtain the energy spectrum for Eqs.(\ref{eq:3dh}) and (\ref{eq:1dh});
\begin{eqnarray}
{\cal E}_{n,\zeta} &=&
\left\{
\begin{array}{ll}
\lambda_{\zeta}\sqrt{1+\frac{2eBn}{\lambda_{\zeta}^{2}}} & n=1,2....\\
\lambda_{\zeta=+} & n=0,
\end{array}
\right.
\label{eq:energy}
\end{eqnarray}  
where $\lambda_\zeta$ is asymmetric respect to $\lambda_\zeta=0$ if complex conjugate symmetry is broken, $M({\bf x})\not=M({\bf x})^*$\cite{Tatsumi:2014wka}.

These results can be easily generalized to the case of $N_f=2$. 
Assuming that the ground state is the charge eigenstate, 
\begin{eqnarray}
M = -2G[\langle\overline{\psi}\psi\rangle+i\langle\overline{\psi}i\gamma^{5}\tau^{3}\psi\rangle]\\
\langle\overline{\psi}i\gamma^{5}\tau^{1}\psi\rangle= \langle\overline{\psi}i\gamma^{5}\tau^{2}\psi\rangle=0,
\end{eqnarray}
the NJL Lagrangian with three colors and two flavors is  written as 
\begin{eqnarray}
\nonumber L &=& \overline{\psi}(i\gamma^{\mu}D_{\mu}-M\frac{1+\tau^3\gamma^{5}}{2}-M^{*}\frac{1-\tau^3\gamma^{5}}{2})\psi- \frac{\mid M\mid^{2}}{4G},
\end{eqnarray}
which is flavor diagonal, so that we can calculate the energy spectrum for each flavor.

Thermodynamic potential is now written as
\begin{eqnarray}
\Omega[\mu,T,B;\Delta(z)] 
&=& \frac{\left<\left| \Delta (z)\right|^2 \right>}{4G} 
-TN_c\sum_f\frac{|e_fB|}{2\pi}\sum_{n,\zeta}\sum_{\lambda_\zeta} {\rm ln}\left[ {\rm 2cosh}(\frac{\lambda_{\zeta}\sqrt{1+\frac{2 |e_fB|n}{\lambda_{\zeta}^2}}-\mu}{2T})   \right] \\
\nonumber &=& \frac{\left<\left| \Delta (z)\right|^2 \right>}{4G} 
-TN_c\sum_f\frac{|e_fB|}{2\pi}\sum_{n,\zeta}\int d\lambda \left[\sum_{\lambda_\zeta} \delta(\lambda - \lambda_{\zeta}) \right]{\rm ln}\left[ {\rm 2cosh}(\frac{\lambda\sqrt{1+\frac{2|e_fB|n}{\lambda^2}}-\mu}{2T})   \right]\\
 &=& \frac{\left<\left| \Delta (z)\right|^2 \right>}{4G} 
-TN_c\sum_f\frac{|e_fB|}{2\pi}\sum_{n,\zeta}\int d\lambda \rho_\zeta(\lambda){\rm ln}\left[ {\rm 2cosh}(\frac{\lambda\sqrt{1+\frac{2|e_fB|n}{\lambda^2}}-\mu}{2T})   \right],
\label{eq:tp1}
\end{eqnarray}
where $\rho_\zeta(\lambda)$ is the density of states, $\rho_\zeta(\lambda)= \sum_{\lambda_\zeta}\delta(\lambda - \lambda_{\zeta})$.

\section{Hybrid Chiral Condensate in the External Magnetic Field \label{sec:hybrid}}

We introduce the hybrid chiral condensate (HCC) which has the properties of both of DCDW and RKC, 
\begin{eqnarray}
M(z)&=& \frac{2m\nu}{1+\sqrt{\nu}}{\rm sn}\left(\frac{2mz}{1+\sqrt{\nu}};\nu \right)e^{iqz},
\end{eqnarray}
and is characterized by three parameters; $m,q,\nu$. 
It is reduced to the pure DCDW in one limit, $\nu\rightarrow 1$, while to the pure RKC in the other limit, $q\rightarrow 0$.
Thus HCC is the minimum configuration which includes both of DCDW and RKC.
Note that HCC is simply given by the product of the two types of the condensate,
but it satisfies the BdG equation within the NJL$_2$ model.
In Refs. \cite{Basar:2008ki,Basar:2009fg}, Basar et al. have found the general form of the condensate in $1+1$ dimensions, 
\begin{equation}
M(z)=-m\, e^{i q z}\, A\,\frac{\sigma(m A\, z+i{\bf K}^\prime -i\theta/2)}
{\sigma(m A\, z+i{\bf K}^\prime)\sigma(i\theta/2)}
\,\exp\left[im A\, z \left(-i\,\zeta(i\theta/2)+i\,{\rm ns}(i\theta/2)\right)+i\,\theta \eta_3/2\right]
 \quad 
 \end{equation}
characterized by four parameters; $m,q,\nu,\theta$,
where $A=A(\theta,\nu)=-2i{\rm sc}(i\theta/4;\nu){\rm nd}(i\theta/4;\nu)$, and $\sigma$ and $\zeta$ are Weierstrass sigma and zeta functions, and $\eta_3= \zeta(i{\rm K}')$. 
When $\theta=2{\bf K}(\nu)$, this condensate becomes HCC. 
It can be easily seen that the energy spectrum in the HCC phase $\lambda_\zeta$ is uniformly shifted by $\zeta q/2$ from the RKC one, $\lambda_\zeta\rightarrow \lambda_\zeta-\zeta q/2$.  
 Accordingly, the density of states is given as
\begin{eqnarray}
\rho_{\zeta}(\lambda) &=&\rho_{\rm RKC}({\lambda-\zeta q/2})\nonumber\\
&=&\frac{1}{\pi}\frac{|(\lambda-\zeta q/2)^2 + m^2c|}{\sqrt{((\lambda-\zeta q/2)^2-m^2)((\lambda-\zeta q/2)^2-m^2{\nu'})}} ,
\label{eq:density}
\end{eqnarray}
 by using the density of states for RKC, $\rho_{\rm RKC}(\lambda)$ given in \cite{Basar:2008ki}, where  
 $c=(1-\nu-2{\bf E(\nu)/K(\nu)})/(1+\sqrt{\nu})^2$, $\nu'=(1-\sqrt{\nu})^2/(1+\sqrt{\nu})^2$ and ${\bf E(\nu),K(\nu)}$ are the complete elliptic integrals.

\begin{figure}[htb]
\includegraphics[width=10cm]{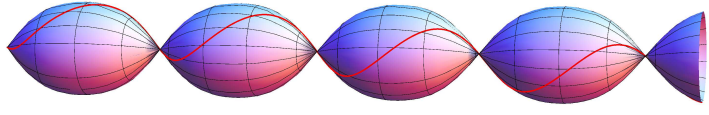}
\caption{ Profile of HCC on the base manifold, which is the direct product of  the horizontal $z$ axis and the vertical chiral circle given by the scalar and pseudo-scalar condensates.}
\label{fig:HCC}
\end{figure}


Putting the density of states (\ref{eq:density}) in Eq.(\ref{eq:tp1}), we have the thermodynamic potential, which is decomposed into the vacuum contribution, the medium contribution and the thermal contribution:
\begin{eqnarray}
\label{eq:potential}
{\Omega}[\mu,T,B;m,\nu,q] &=& \frac{m^2}{4G}\left( 1-\frac{{\bf E}(\nu)}{{\bf K}(\nu)}\right) +\Omega _{\rm vac} + \Omega _{\mu} +\Omega _{T} \\
\Omega _{\rm vac} &=&-\frac{1}{2}N_c \sum_{f} \frac{|e_fB|}{2\pi}\sum_{n,\zeta}\int d\lambda \rho_\zeta(\lambda)|\lambda\sqrt{1+\frac{2|e_fB|n}{\lambda^2}}| \\
\Omega _{\mu}&=&-\frac{1}{2}N_c \sum_{f} \frac{|e_fB|}{2\pi}\sum_{n,\zeta}\int d\lambda \rho_\zeta(\lambda)
 \left[ |\lambda\sqrt{1+\frac{2|e_fB|n}{\lambda^2}}-\mu|-|\lambda\sqrt{1+\frac{2|e_fB|n}{\lambda^2}}| \right]\\
 \Omega _{T} &=&-TN_c \sum_{f} \frac{|e_fB|}{2\pi}\sum_{n,\zeta}\int d\lambda \rho_\zeta(\lambda){\rm ln}\left[ 1+ {\rm exp}(-\frac{|\lambda\sqrt{1+\frac{2|e_fB|n}{\lambda^2}}-\mu|}{T})   \right]
\end{eqnarray}
Here the vacuum contribution $ \Omega _{\rm vac}$ is divergent, so that we use the proper time regularization. 
\begin{eqnarray}
\nonumber \Omega _{\rm vac} &=& N_c \sum_{f,\zeta} \frac{|e_fB|}{16\pi^{3/2}}\int_{1}^{\infty}\frac{d\tau}{\tau^{3/2}}{\rm coth}(\tau|e_fB|)
\int d\lambda\rho_{\zeta}(\lambda){\rm exp}\left(-\tau \lambda^2\right)
\end{eqnarray}

\section{Spectral asymmetry and anomalous quark number density\label{sec:sa}}
The fermion number is given by
\begin{eqnarray}
\label{eq:number}
N_B &=&-\frac{1}{2}\eta_H + VN_c \sum_{f} \frac{|e_fB|}{2\pi}\sum_{n,\zeta}\int d\lambda \rho_\zeta(\lambda)\left[\frac{\theta({\cal E})}{1+e^{({\cal E}-\mu)/T}}+\frac{\theta({\cal -E})}{1+e^{-({\cal E}-\mu)/T}}\right],
\end{eqnarray}
where the first term is the fermion number from {\it spectral asymmetry} \cite{Niemi:1984vz} characterized by the Atiyah-Patodi-Singer $\eta$-invariant \cite{Niemi:1984vz,Atiyah1973} which is written as
\begin{eqnarray}
\nonumber \eta_H &=& VN_c \sum_{f}\frac{|e_fB|}{2\pi}\left[ \sum_{\lambda>0}1-\sum_{\lambda<0}1\right]\\
&=& VN_c \sum_{f}\frac{|e_fB|}{2\pi}\lim_{s \to +0} \int_{-\infty}^{\infty}d\lambda \rho_+(\lambda){\rm sign}(\lambda)\left|\lambda\right|^{-s},
\end{eqnarray}
in our case. 
Here, we have used the fact that spectral asymmetry appears only in the spectrum of the lowest Landau level ($n=0$),
and the higher Landau levels ($n \not = 0$) have no contribution to spectral asymmetry. 
Note that the spectrum becomes symmetric without magnetic field.
The second term in Eq.(\ref{eq:number})  counts the number of states for the given $\mu$ with the Fermi-Dirac distribution functions and  is usual number density: 
all  the Landau levels contribute to this term.
Considering quark number density is related to the thermodynamic potential through the thermodynamic relation (\ref{eq:potential}): ${N_B}/{V} = -{\partial\Omega}/{\partial\mu}$,
 we can explicitly verify that this derivative of thermodynamic potential and Eq.(\ref{eq:number}) are equivalent  by using Eq.(\ref{eq:potential}).

\begin{figure}[htb]
\begin{center}
\subfigure[DCDW ($m=0.5,\nu=1,q=0.4$)]{\includegraphics[width=5cm,angle=0]{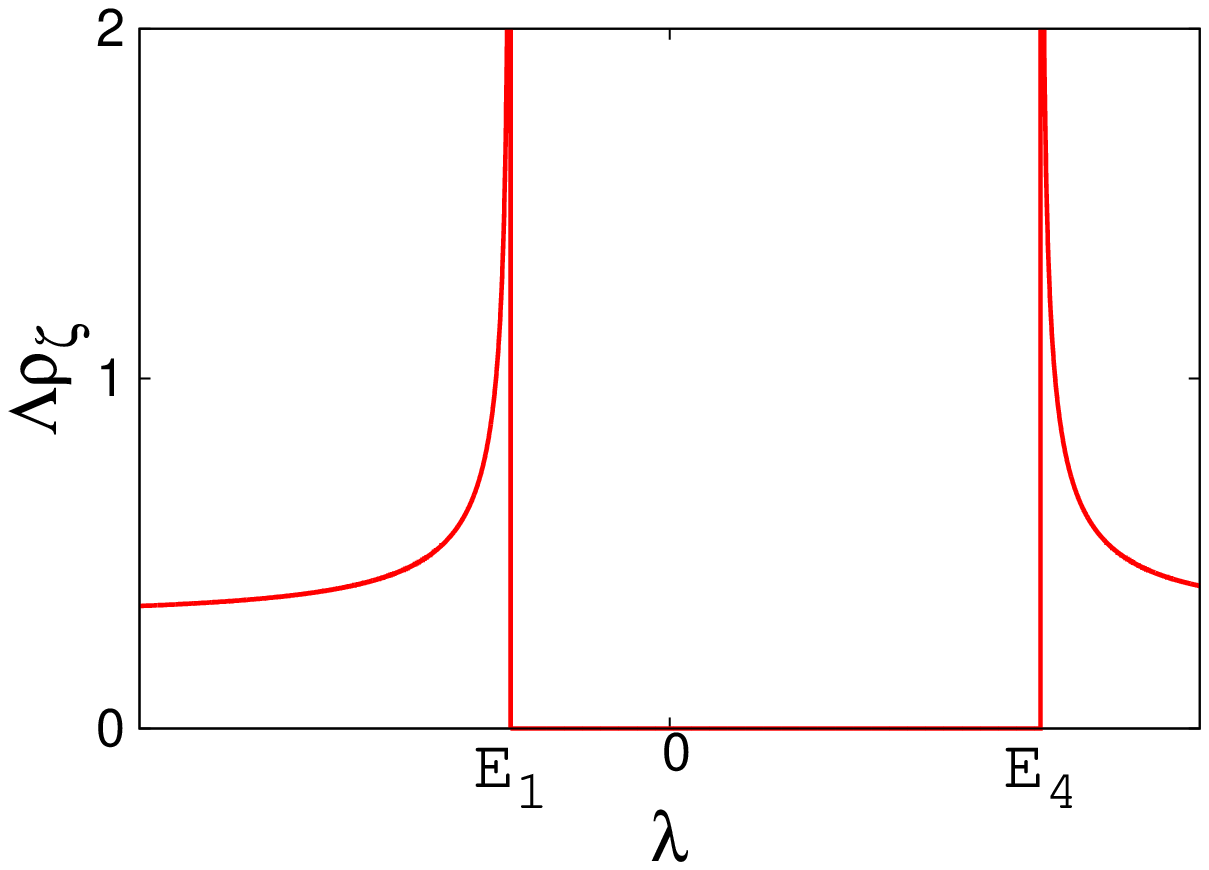}}
\hspace{5mm}
\subfigure[RKC ($m=0.5,\nu=0.3,q=0$)]{\includegraphics[width=5cm,angle=0]{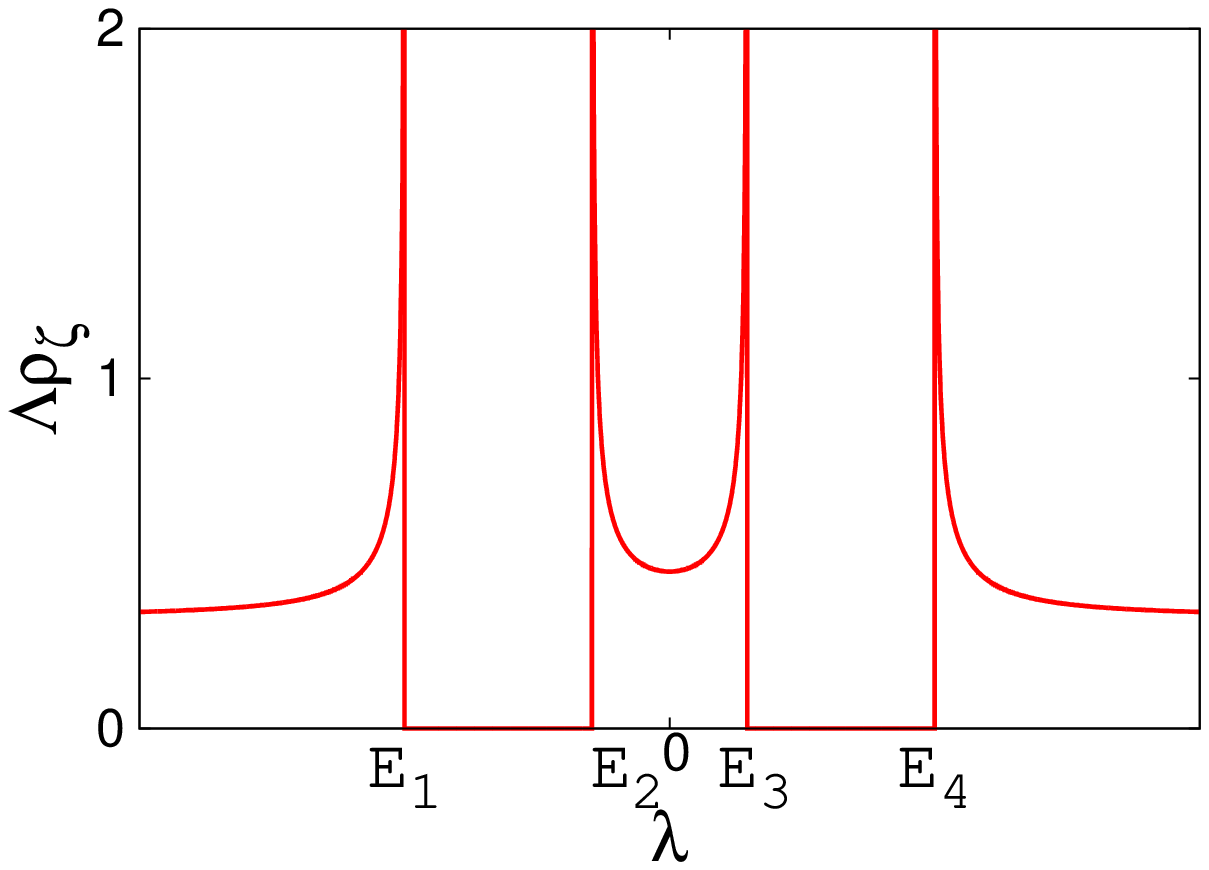}}
\hspace{5mm}
\subfigure[HCC ($m=0.5,\nu=0.3,q=0.4$)]{\includegraphics[width=5cm,angle=0]{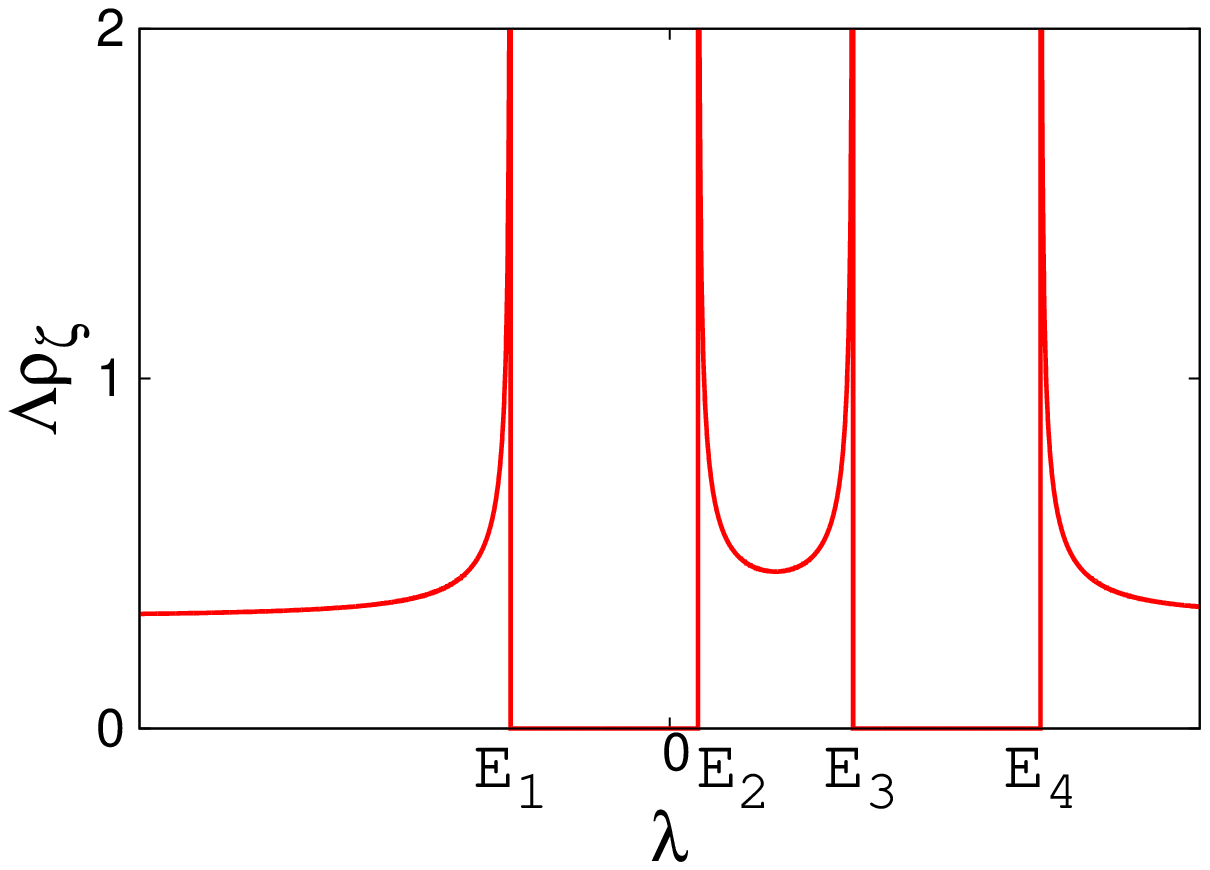}}
   \end{center}
\caption{The behaviour of $\rho_+(\lambda)$.
$E_1=-m+q/2$,~$E_2=-m\nu'+q/2$,~$E_3=m\nu'+q/2$,~$E_4=m+q/2$.
DCDW has the gap between $\lambda=-m+q/2$ and $\lambda=m+q/2$, and has no mid-gap states.
The spectrum of DCDW is asymmetric with respect to 0.
 The spectrum of RKC is symmetric and has the mid-gap states.
 In the case of HCC, spectrum  is asymmetric and has  the mid-gap states.
}
\label{fig:spectrum}
\end{figure}

The density of states of the lowest Landau level (LLL) is schematically shown in Fig. \ref{fig:spectrum}.
When the Hamiltonian is symmetric for complex conjugation operation $M({\bf x})\to M({\bf x})^*$, the spectrum is symmetric. 
So in the case of RKC or homogeneous condensate, number density becomes usual number density.
In the case of HCC or DCDW, the contribution of  $\eta$-invariant is nonzero in the presence of the magnetic field.
The  $\eta$-invariant has been evaluated to give 
\begin{equation}
\eta_H=-VN_c \sum_{f}\frac{|e_fB|}{2\pi}\frac{q}{\pi}
\label{eta-DCDW}
\end{equation}
for DCDW in the case of $-m+q/2<0$ \cite{Tatsumi:2014wka}.
It has a topological origin, and is equal to the expression given by chiral anomaly \cite{Son:2007ny} (see Appendix \ref{sec:spectral2}).
It is straightforward to evaluate the $\eta$-invariant for the case of HCC (Appendix \ref{etainv}),
\begin{equation}
\eta_H=-VN_c \sum_{f}\frac{|e_fB|}{2\pi}\left(\frac{q}{\pi}-N_{\rm midgap}\right),
\end{equation}
for the case of $-m+q/2<0$, where the second term is the contribution from the mid-gap states. 
In particular, for $m > q/2 > m\nu'$, it equals to the number of nodes of HCC, $N_{\rm nodes}=m/[(1+\sqrt{\nu}){\bf K}(\nu)]$, independent of $q$.
Note that this is the same form as in the DCDW phase, except the number of nodes of the condensates.
 For the general case, the $\eta$-invariant can be written as 
$
\eta_H=VN_c \sum_{f}\frac{|e_fB|}{2\pi}\left[ -\frac{q}{\pi} +\frac{2m}{\pi}{\rm Re}\left({\bf F}(q/2;\nu^\prime) +c{\bf F}(q/2;\nu^\prime)-{\bf E}(q/2;\nu^\prime) \right)\right],
$
where ${\bf F}(x;\nu^\prime)$ and ${\bf E}(x;\nu^\prime)$ are the incomplete elliptic integrals of first and second kind. Note that $\eta_H$ is reduced to the DCDW one (\ref{eta-DCDW}) 
as $q\rightarrow 0$, where the energy spectrum is reduced to the one of RKC and symmetric about zero.

For low chemical potential, $m+q/2>\mu$, usual number density is zero, and the inhomogeneous phase is forbidden by the Lorentz symmetry of the vacuum.
On the other hand, spectral asymmetry gives nonzero number density, and the appearance of the inhomogeneous phase is allowed in the presence of the magnetic field.
The contribution of spectral asymmetry is taken in the thermodynamic potential as the term, $\mu \eta_H /2$. 
Since this term includes the linear term in $q$ and the order parameters are determined by the stationary conditions for the thermodynamical potential,  
 $q=0$ is never the optimal point. In other words the RKC phase itself does not appear in the QCD phase diagram in the magnetic field.

In \cite{Tatsumi:2014wka}, spectral asymmetry has been also calculated using the derivative expansion and $\eta_H=N_c\sum_f |e_f|{\bf B\cdot q}/2$ has been obtained for DCDW.
Consequently the ${\bf q}\cdot{\bf B}$ term should appear in the thermodynamic potential by way of the thermodynamic relation, and $\bf q$ is favored to be parallel to $\bf B$.
For HCC, the derivative expansion can't be directly applied because the condensate has nodes and the premise that the amplitude is much larger than the wavevector breaks down.
However, we can manage to evaluate the $\eta$-invariant by separating the small nodal region to find $\eta_H=N_c\sum_f |e_f|{\bf B\cdot q}/2+N_{\rm nodes}$ for $m\nu'<q/2$,  which suggests 
$\bf q$ is most favored to be parallel to $\bf B$ as well in the HCC phase.

\section{Phase diagram\label{sec:phase}}
For obtaining the phase diagram, we numerically search the minima of the thermodynamic potential with respect to the order parameters; $m$, $q$ and $\nu$  for given values of the magnetic field $B$ and chemical potential $\mu$.
In this paper, we show the phase diagram at zero temperature.
We use $G\Lambda^2 = 6.35$ and $\Lambda = 660$MeV which reproduce $f_\pi = 93$MeV and the constituent quark mass $\simeq 330$MeV in the vacuum.

 \subsection{RKC phase in the magnetic field}
First, we consider the effect of the magnetic field on the RKC phase.
Without the magnetic field, the RKC phase is energetically more favorable than the DCDW phase in the framework of the NJL model \cite{Nickel:2009wj}.
For RKC, the order parameter $M(z)$ is real and there appears no spectral asymmetry.

\begin{figure}[htb]
\begin{center}
\includegraphics[width=9cm,angle=0]{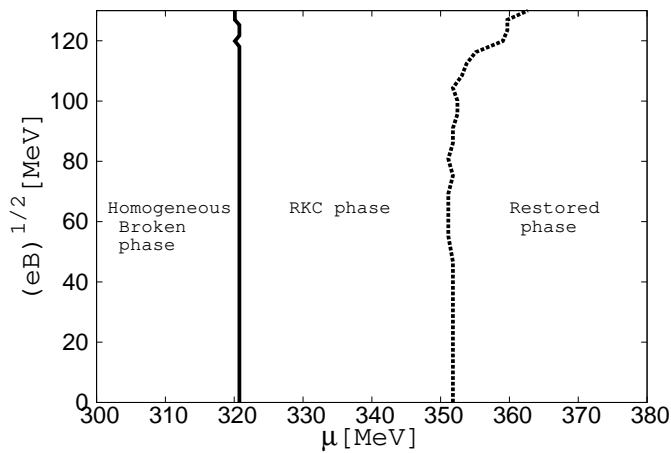}
   \end{center}
\caption{Phase diagram for the RKC phase.
}
\label{fig:rkcphase}
\end{figure}

The phase diagram of the RKC phase is essentially unchanged, while ,as we shall see later, that of the DCDW phase is significantly changed in the presence of the magnetic field. 
Since the chiral condensate is neutral, one may expect that there is little effect of the magnetic field on the inhomogeneous chiral phase. 
Although it holds for the RKC phase, some anomalous effect coming from spectral asymmetry plays an important role in the DCDW phase. 
Figure \ref{fig:rkcphase} shows the phase diagram of the RKC phase in the presence of the magnetic field.
We can see some oscillation of the  phase boundary with respect to the magnetic field, which comes from the Landau quantization and related to the de Haas-van Alphen effect \cite{deHaas1936,Ebert:1999ht}.
This oscillation is also observed in the case of the homogeneous chiral condensate within the NJL model \cite{Inagaki:2003yi}.

\begin{figure}[htb]
\begin{center}
\subfigure[$\sqrt{eB}=60 {\rm MeV}$]{\includegraphics[width=7cm,angle=0]{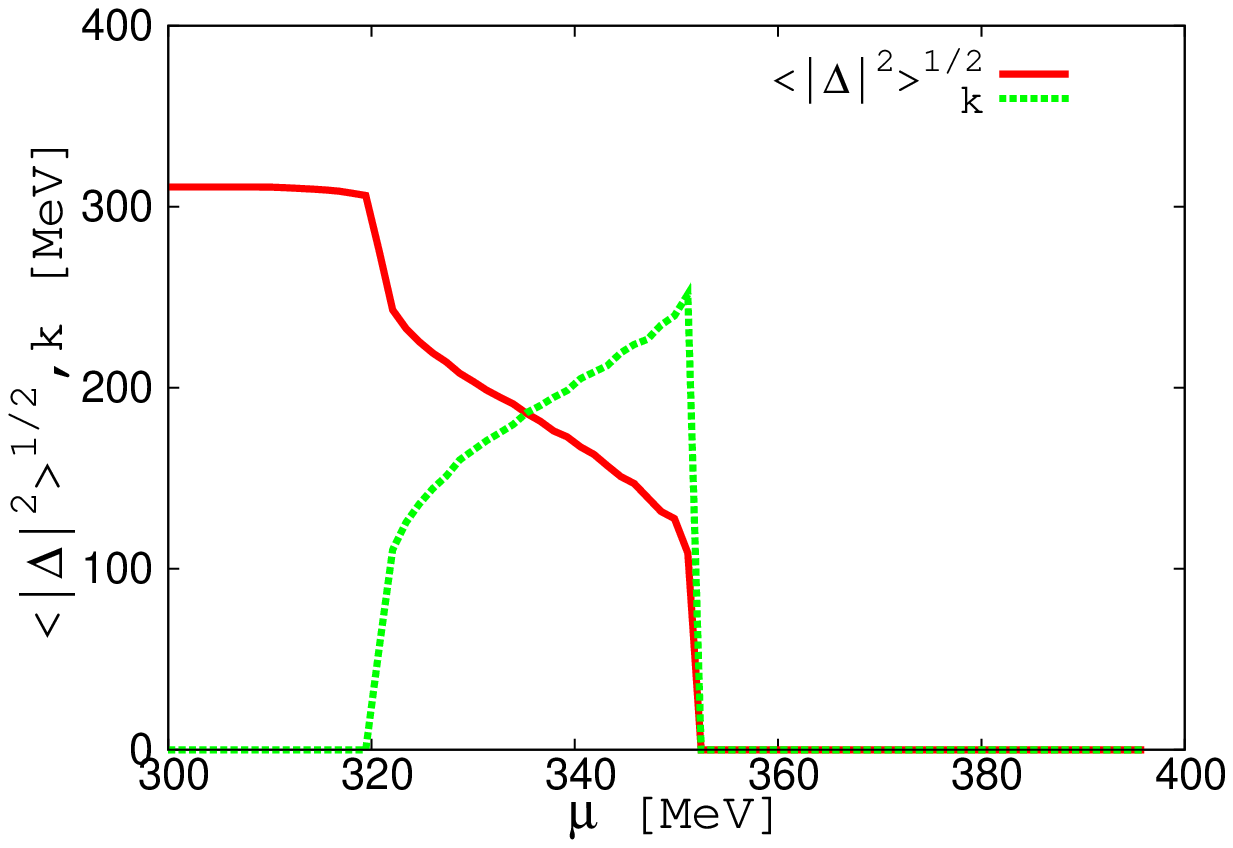}}
\hspace{9mm}
\subfigure[$\sqrt{eB}= 120{\rm MeV}$]{\includegraphics[width=7cm,angle=0]{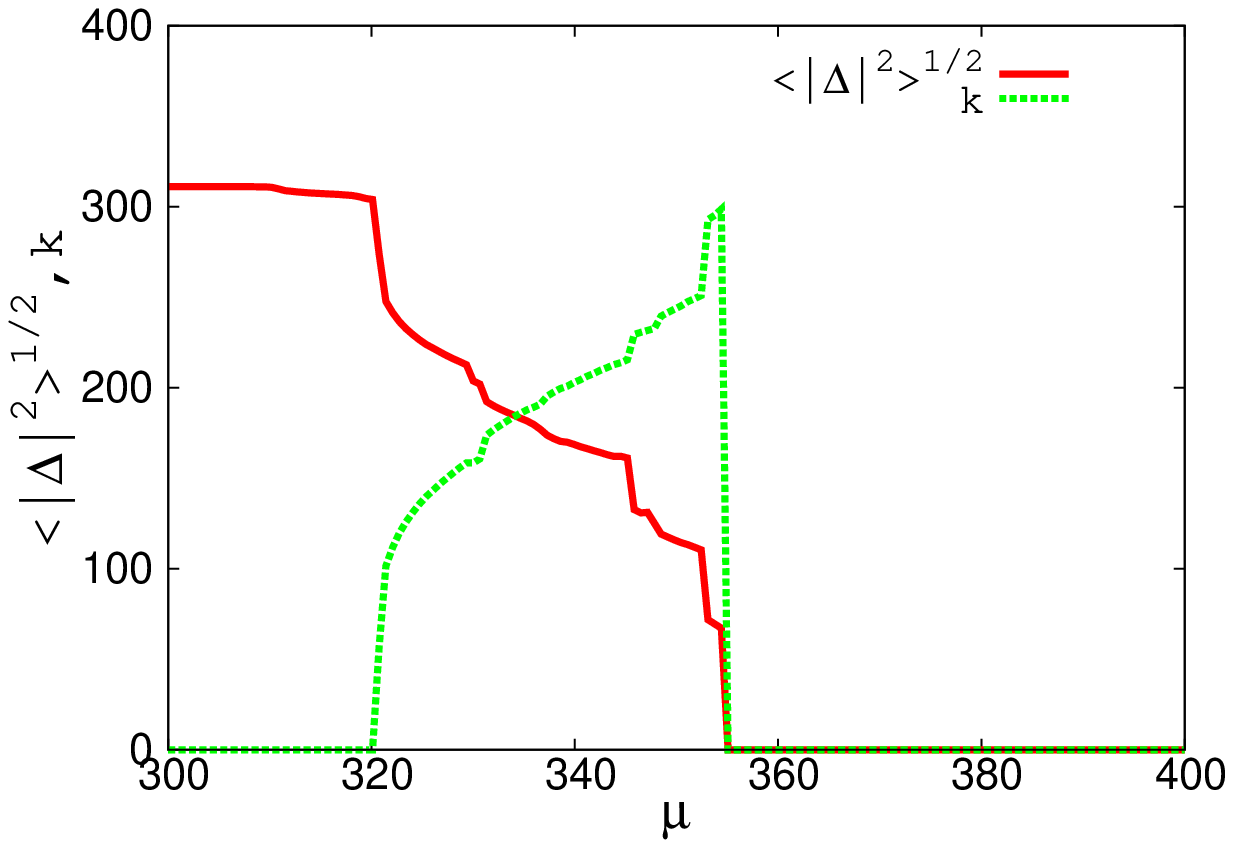}}
\vspace{12mm}
   \end{center}
\caption{Order parameters as functions of chemical potential in different magnetic field,
where $k$ is the wavevector of RKC : $k=2m/[(1+\sqrt{\nu}){\bf K}(\nu)]$
}
\label{fig:rkcorder}
\end{figure}

Figure \ref{fig:rkcorder} shows the order parameters as functions of chemical potential in different magnetic fields.
At  $\sqrt{eB} = 60 {\rm MeV}$, the order parameters behave like continuous functions of the chemical potential, and are very similar to those in the absence of the magnetic field.
At  $\sqrt{eB} = 120{\rm MeV}$, the order parameters exhibit some discontinuous jumps, since the thermodynamic potential  has some local minima as a function of $m$ and $\nu$ in this region.

\subsection{HCC phase in the magnetic field}
Here, we consider the phase diagram by introducing the HCC, which includes both  features of DCDW and RKC,
and use the following approximation instead of fully evaluating the thermodynamic potential.
When the magnetic field is much weak, the approximation
\begin{eqnarray}\label{eq:approx}
\Omega(B) &\simeq& \Omega(B=0)+eB\Omega^{(1)}
\end{eqnarray}
is valid, where the first order correction is written in
\begin{eqnarray}
eB\Omega^{(1)} &=&1/2\left( \Omega_{LLL,q}-\Omega_{LLL,-q} \right)
\end{eqnarray}
(See Appendix:\ref{sec:expansion}).
This term is an odd function of $q$, and vanishes at $q=0$,
so that this term does not appear in the thermodynamic potential for the RKC phase.
Only LLL contributes to $\Omega^{(1)}$, while the higher Landau levels ($n\neq 0$) and LLL contribute to the second and higher order terms.
We checked the validity of this expansion for some $eB$ by comparing the numerical results with Eq.~(\ref{eq:approx}) and the full expression Eq.~(\ref{eq:potential}).
Consequently we have found that the phase structure is almost unchanged for $\sqrt{eB} < 0.2\Lambda \simeq 120{\rm MeV}$.

The magnetic properties of the DCDW phase in the external magnetic field has been studied by Frolov et al. \cite{Frolov:2010wn}.
They have shown that the DCDW phase is always favorable than that of the homogeneous chiral condensate in the presence of $B$.
As is already stated in the previous section, it is shown that the mechanism of superiority of the DCDW phase is related to spectral asymmetry of LLL states.

\begin{figure}[htb]
\begin{center}
\includegraphics[width=12cm,angle=0]{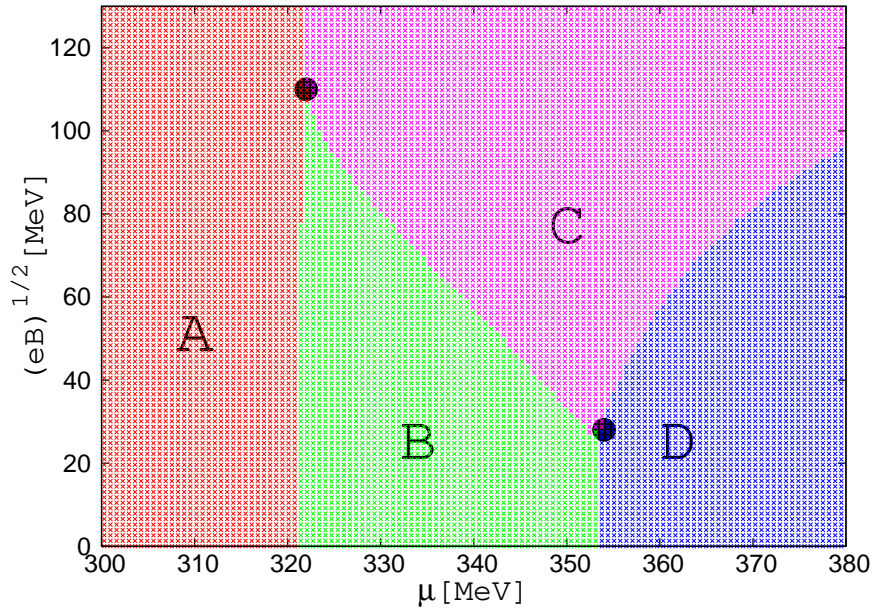}\\
\vspace{12mm}
   \end{center}
\caption{Phase diagram at T=0.
A: Weak DCDW phase
B: HCC phase
C: Strong DCDW phase
D: Chiral-restored phase.
When $eB \to 0$ , the weak DCDW phase becomes the homogeneously chiral-broken phase smoothly, and the HCC phase becomes the RKC phase.
The filled circles represent triple points.}
\label{fig:phase}
\end{figure}

The phase diagram is shown in Fig.\ref{fig:phase}.
In this figure, A denotes the weak DCDW phase, B the HCC phase ,C the strong DCDW phase, and D the chiral-restored phase.
The triple points appear at $(\mu,\sqrt{eB}) \sim (320{\rm MeV},110{\rm MeV})$ and $(\mu,\sqrt{eB}) \sim (350{\rm MeV},{\rm 30MeV})$.
For the limit $eB \to 0$ , the weak DCDW phase is reduced to the homogeneously chiral-broken phase and the HCC phase to the RKC phase.
In the $eB \not = 0$ region, the order parameter is always finite, and 
there is no homogeneously chiral-broken phase nor the RKC phase if the magnetic field has nonzero strength.
The phase boundary between the chiral-broken and restored phases moves to higher $\mu$ as the magnetic field becomes stronger.
Thus the magnetic field expands the chiral-broken phase mainly due to the phase degree of freedom.

\begin{figure}[htb]
\begin{center}
\subfigure[$\sqrt{eB}=0,\mu=0.48\Lambda$]{\includegraphics[width=5cm,angle=0]{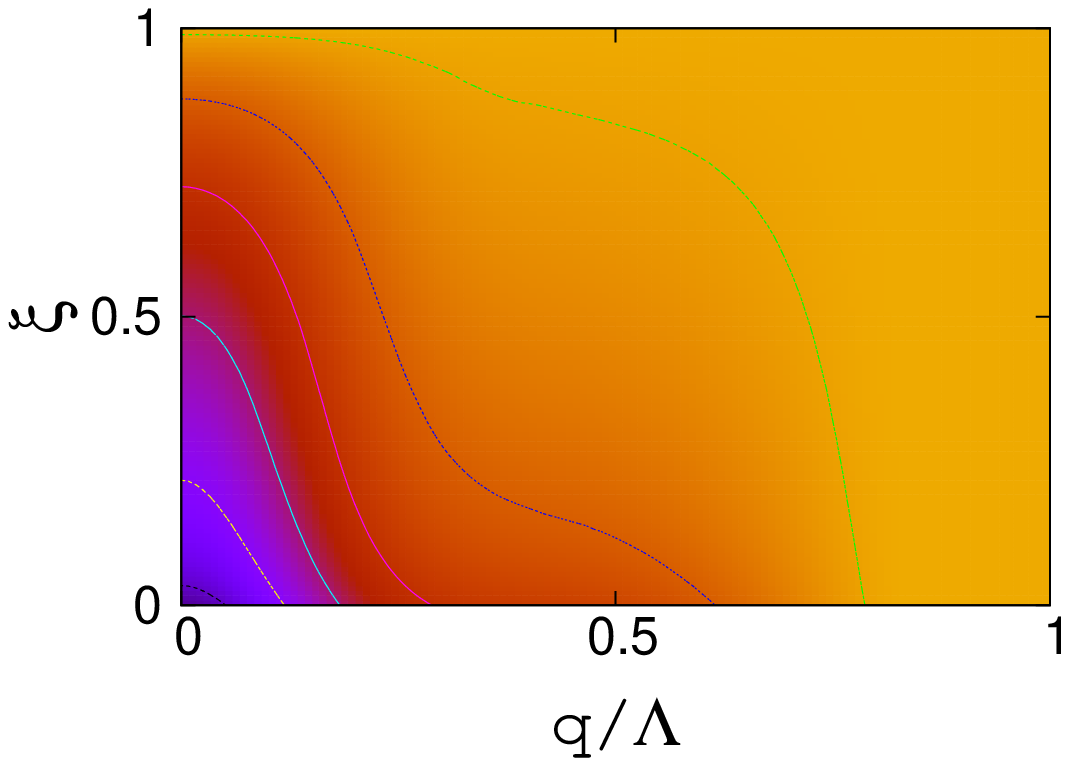}}
\subfigure[$\sqrt{eB}=0,\mu=0.485\Lambda$]{\includegraphics[width=5cm,angle=0]{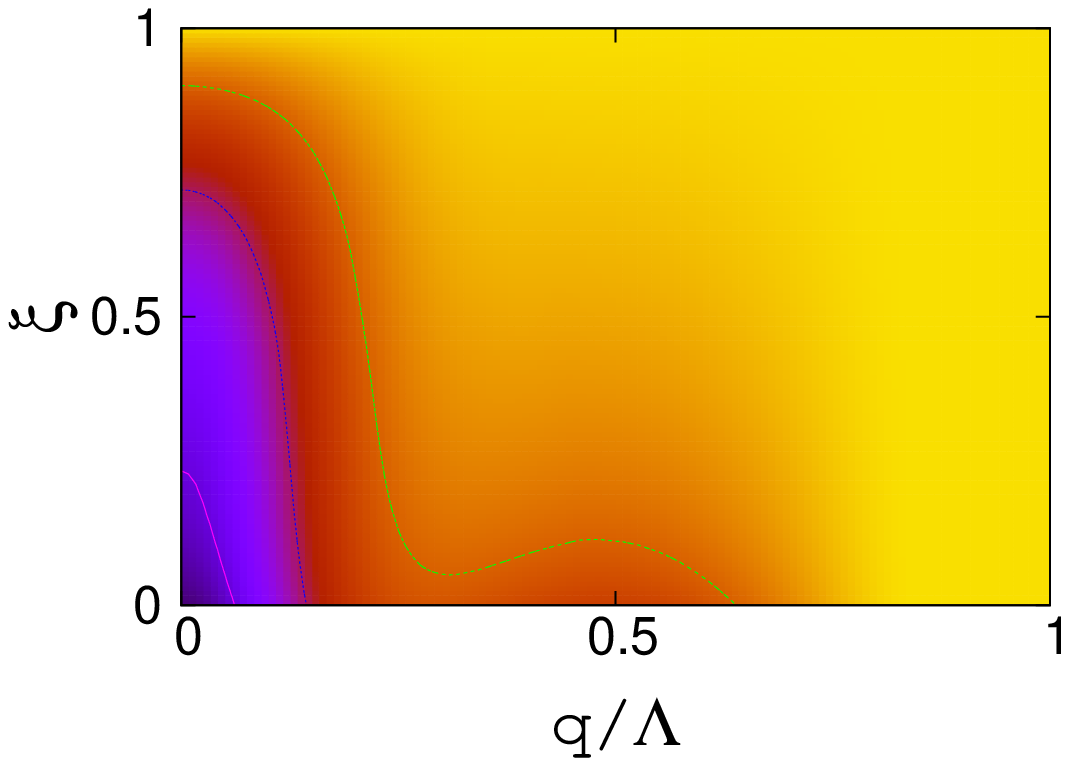}}
\subfigure[$\sqrt{eB}=0,\mu=0.49\Lambda$]{\includegraphics[width=5cm,angle=0]{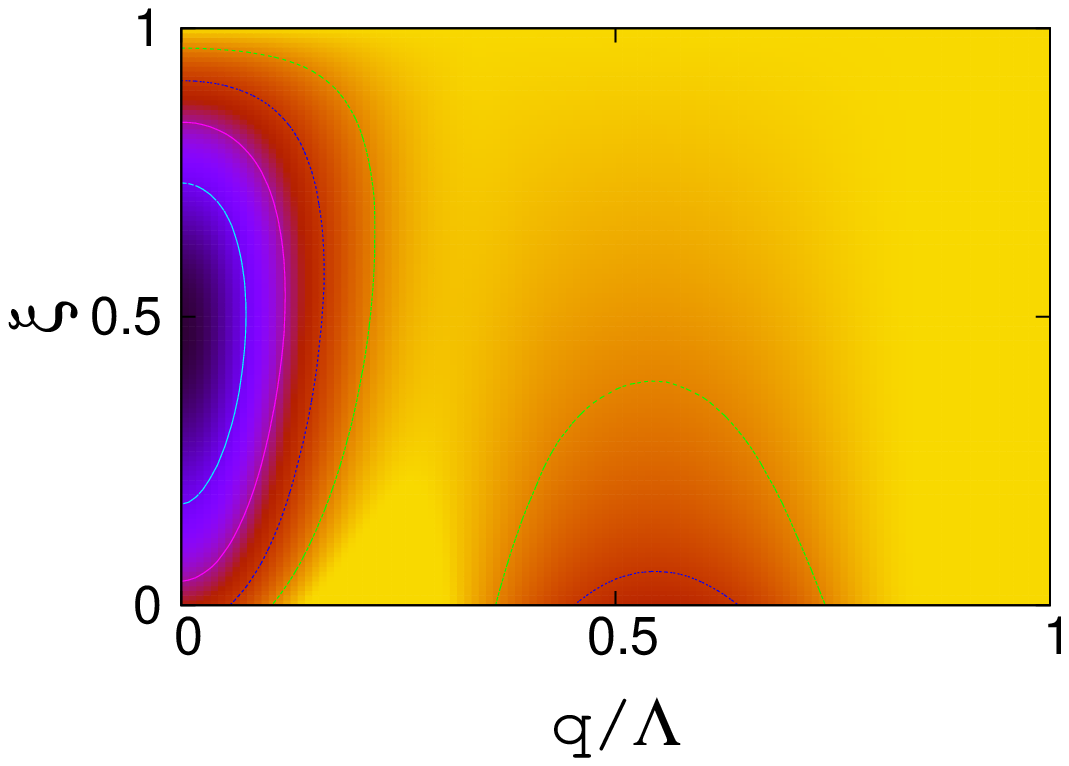}}
\subfigure[$\sqrt{eB}=0,\mu=0.5\Lambda$]{\includegraphics[width=5cm,angle=0]{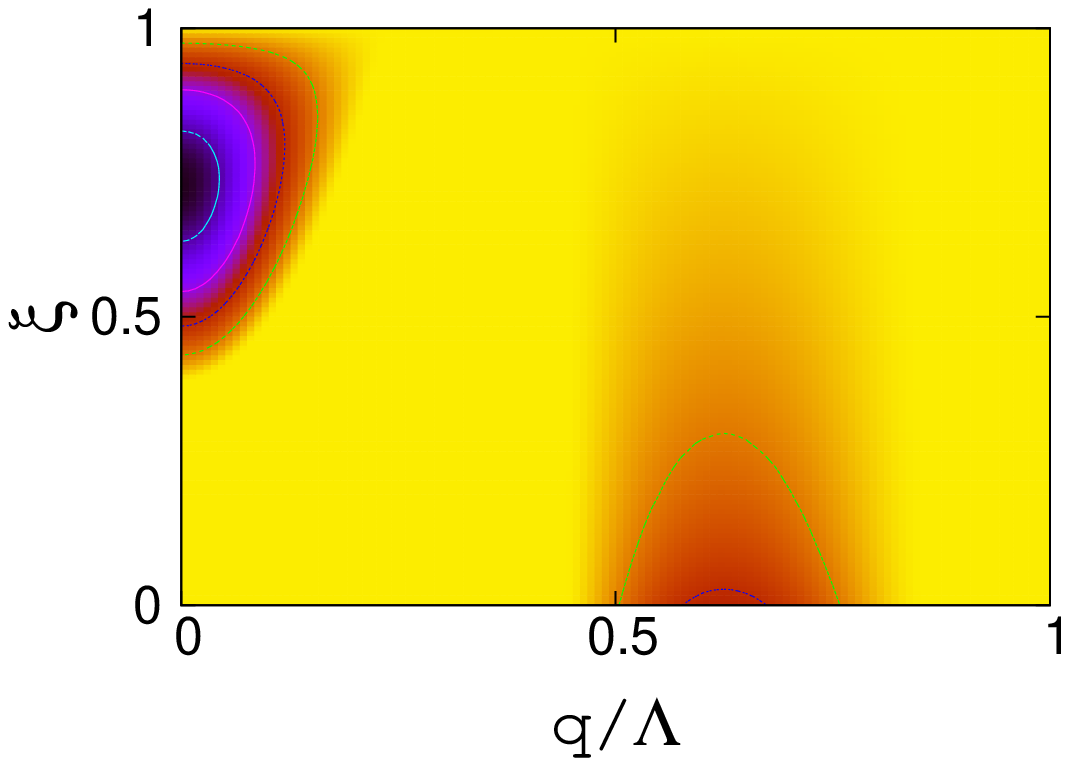}}
\subfigure[$\sqrt{eB}=0.18\Lambda,\mu=0.48\Lambda$]{\includegraphics[width=5cm,angle=0]{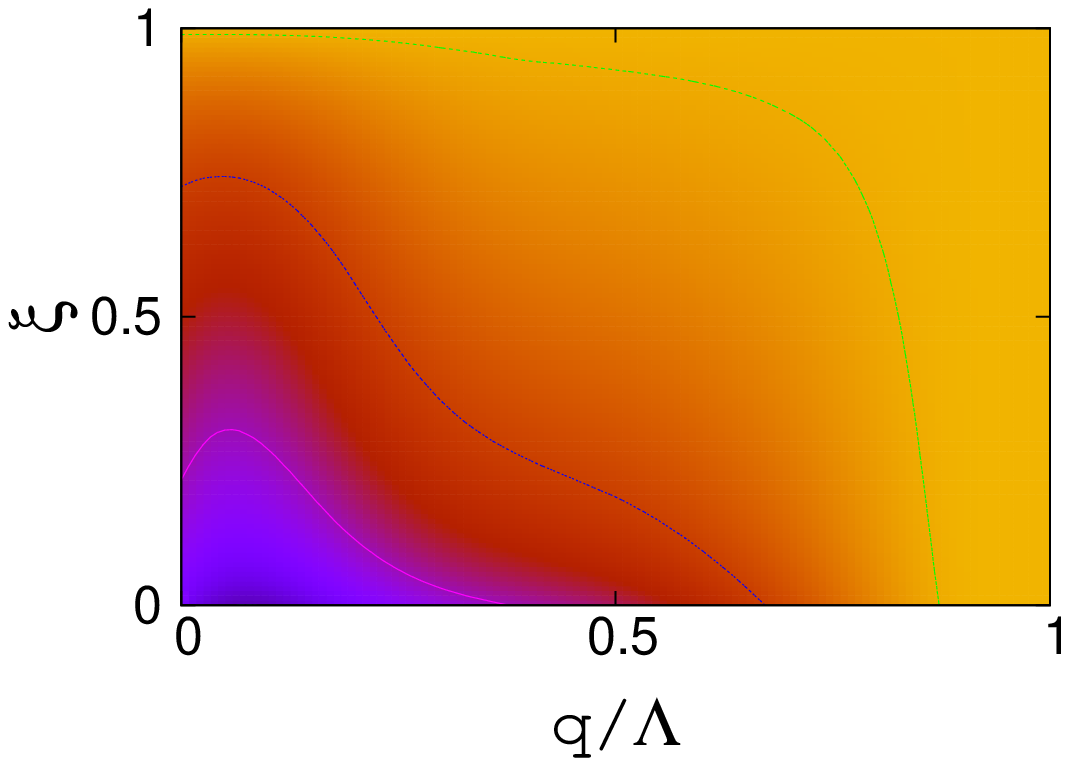}}
\subfigure[$\sqrt{eB}=0.18\Lambda,\mu=0.485\Lambda$]{\includegraphics[width=5cm,angle=0]{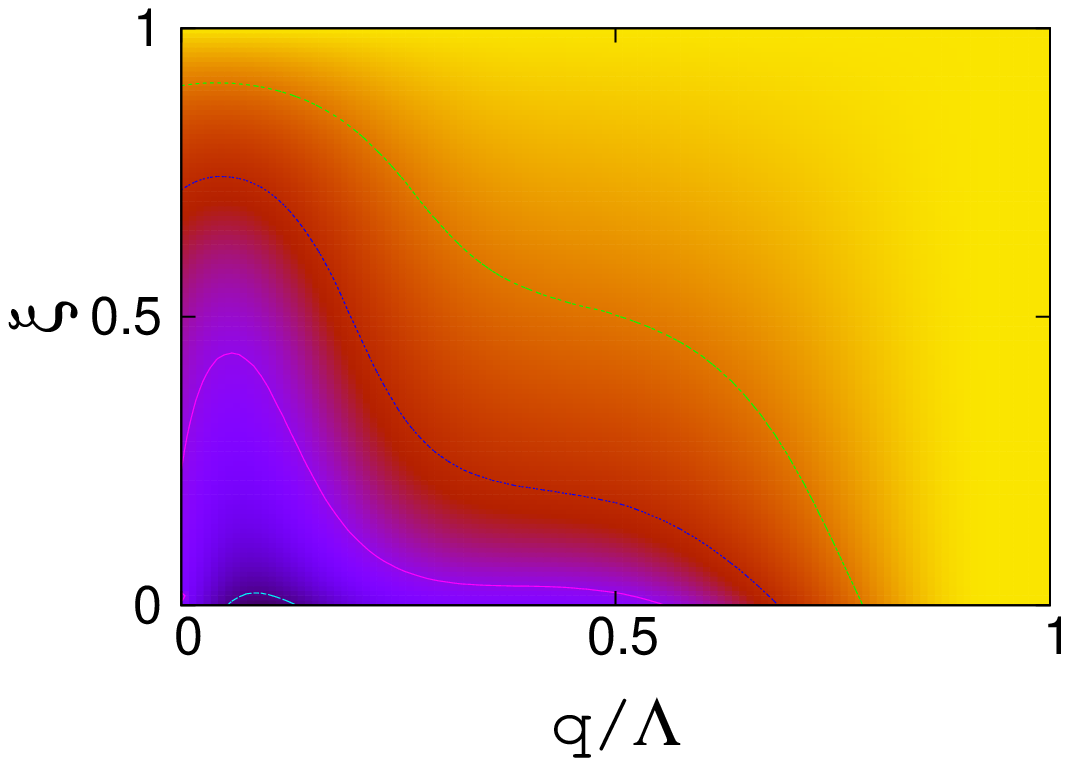}}
\subfigure[$\sqrt{eB}=0.18\Lambda,\mu=0.49\Lambda$]{\includegraphics[width=5cm,angle=0]{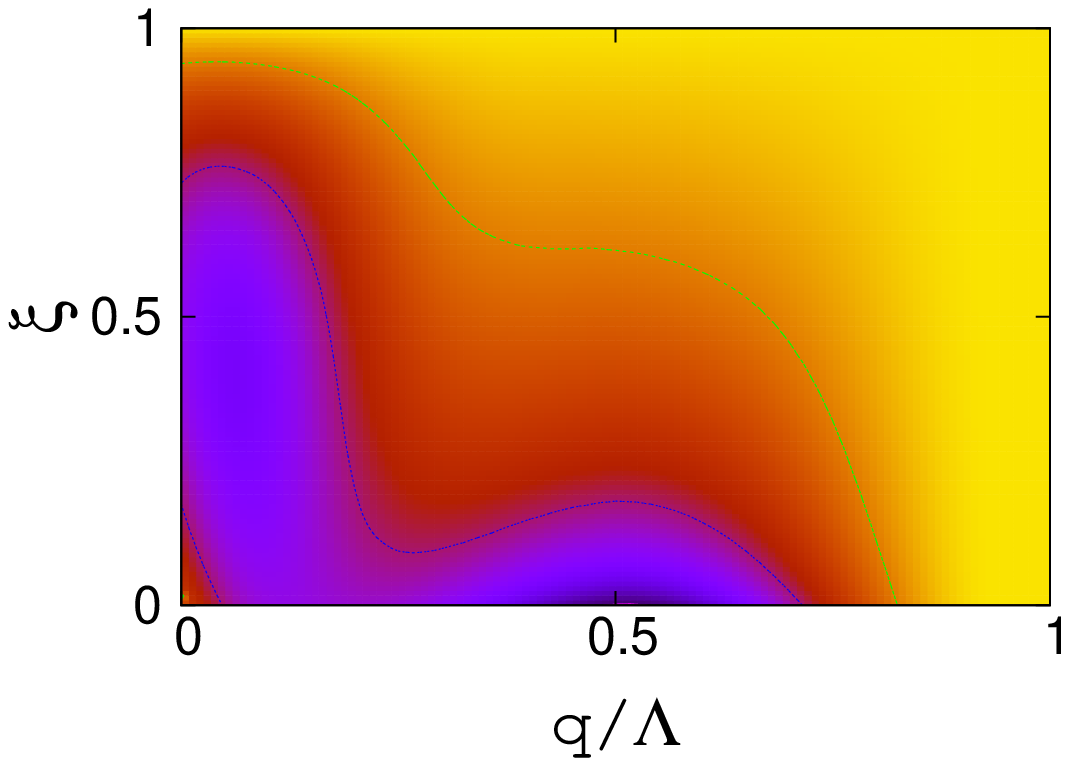}}
\subfigure[$\sqrt{eB}=0.18\Lambda,\mu=0.5\Lambda$]{\includegraphics[width=5cm,angle=0]{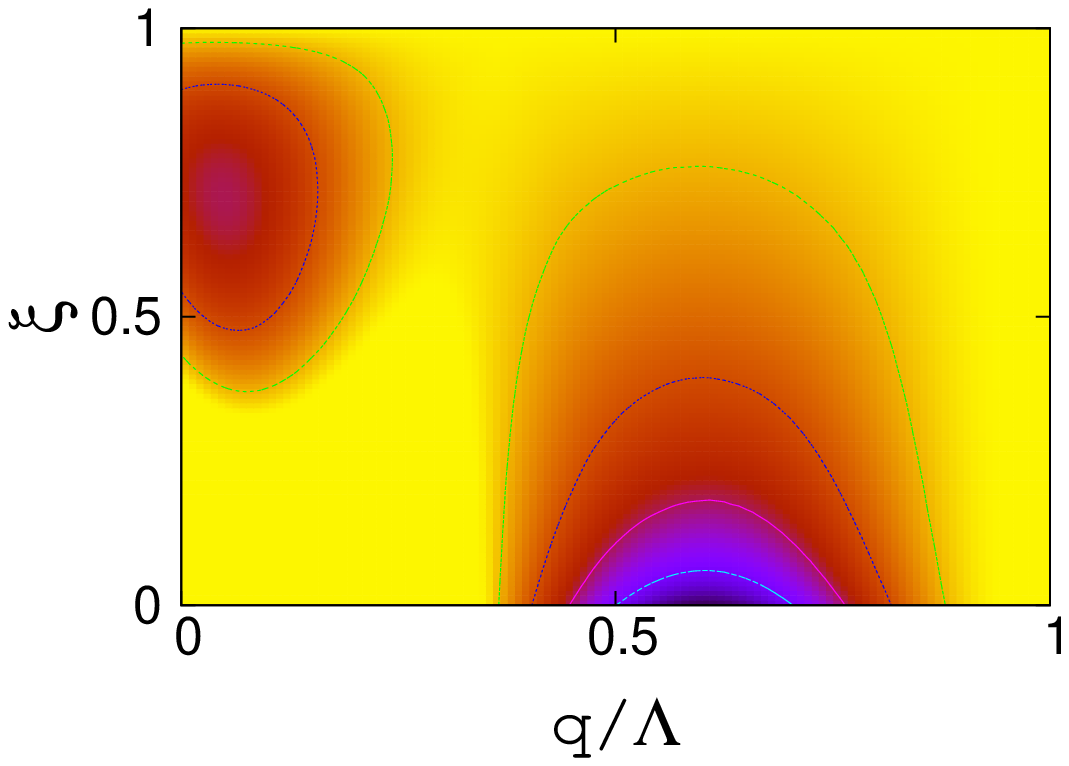}}
   \end{center}
\caption{Energy surface of thermodynamic potential at different $eB$ and different $\mu$}
\label{fig:surface}
\end{figure}

Figure \ref{fig:surface} shows the energy surface of the thermodynamic potential in the $\xi-q$ plane,
where $m$  is set to be the optimal value for given $\xi,q$ with $\xi = 16^{(1-1/\nu)}$.
Without the magnetic field, there are two local minima corresponding to DCDW and RKC in the energy surface of the thermodynamic potential;
RKC is energetically more favored than DCDW.
The minimum with the homogeneous chiral condensate is smoothly changed to that of RKC, so this phase transition is  of the second order.
There is a competition between RKC and DCDW, and there appears no phase in which both of phase and amplitude modulations are large.
Once turning on  the magnetic field,  both minima of RKC and DCDW move to the larger $q$ direction, which is caused by spectral asymmetry.

It has been discussed that the mechanism of emergence of  the inhomogeneous chiral phase is the Fermi surface nesting \cite{Nakano:2004cd}.
Complete nesting is realized in 1+1 dimensions, when the wavenumber of condensate $k_c$ and the Fermi wavenumber $k_F$ have  the relation $2k_F=k_c$.
In 1+3 dimensions, the nesting effect is incomplete, but its reminiscence is left in the DCDW phase \cite{Nakano:2004cd}: $k_c$ is large, $k_c\sim O(2k_F)$. 
The situation is a little changed in the HCC phase.
Using  an approximation $M(z) \simeq m{\rm cos}(kz)e^{iqz}$, we can decompose the order parameter into two different component:
$M(z) = \frac m2(e^{i(q+k)z}+e^{i(q-k)z}) = \frac m2(e^{iq_+z}+e^{iq_-z})$.
The $q_+$ and $q_-$ can not satisfy the nesting relation simultaneously if $k \not = 0$ and $q\not =0$.
Thus HCC can not satisfy the nesting relation, and there is no HCC phase without  the magnetic field.
In the magnetic field, spectral asymmetry contributes to the emergence of  the HCC phase. 
Note that non-zero value of $q$ is favored by some topological effects in this case, different from the nesting effect.

\begin{figure}[htb]
\begin{center}
\includegraphics[width=7cm,angle=0]{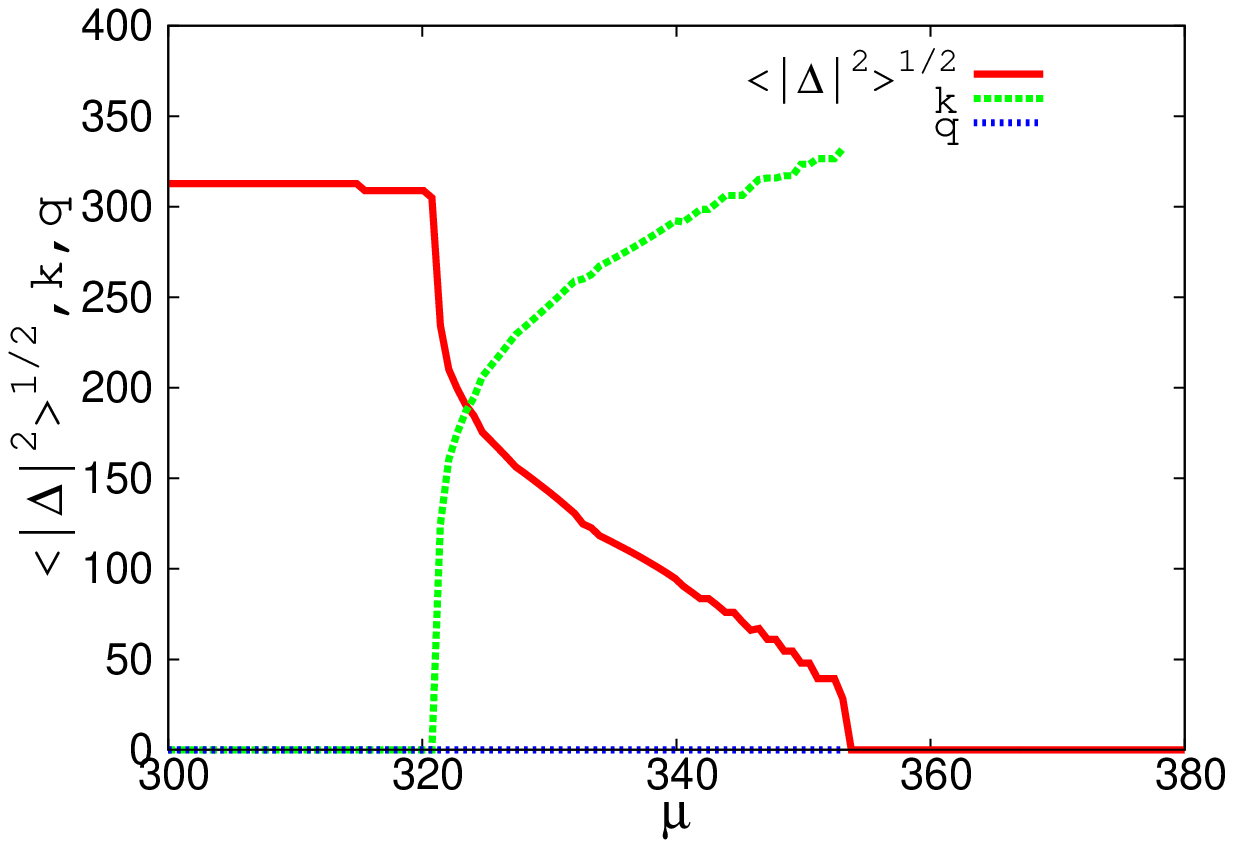}
\hspace{9mm}
\includegraphics[width=7cm,angle=0]{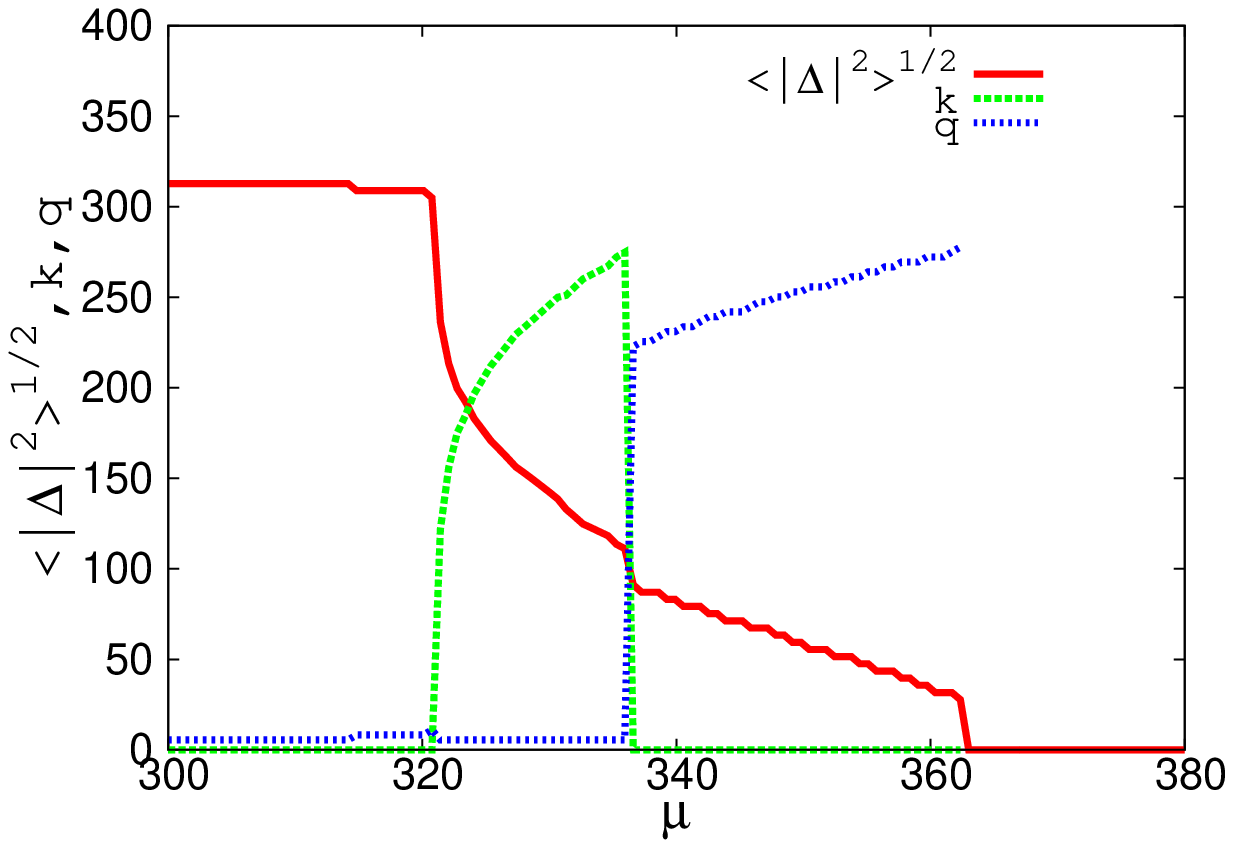}
\hspace{9mm}
\includegraphics[width=7cm,angle=0]{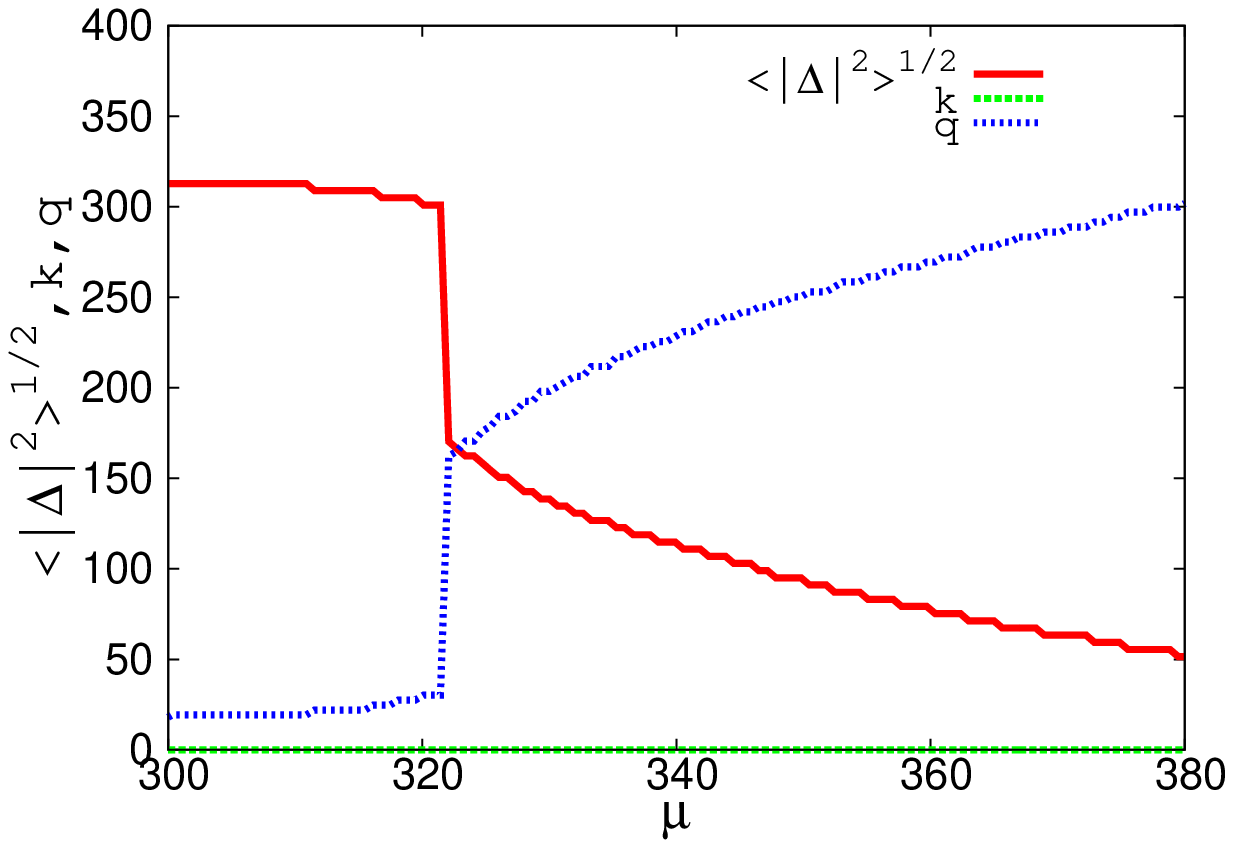}
\vspace{12mm}
   \end{center}
\caption{Order parameters as functions of chemical potential in different magnetic field.
From left to right:
(a) $eB=0$
(b) $\sqrt{eB}=060{\rm MeV}$
(c) $\sqrt{eB}=120{\rm MeV}$
}
\label{fig:order}
\end{figure}

The order parameters are shown as functions of chemical potential in Fig.\ref{fig:order}.
Under no magnetic fields $q$ is zero everywhere, so that there are the homogeneously chiral-broken phase  ($\xi=0$), 
the RKC phase ($\xi\neq 0$), and the chiral-restored phase ($\xi=1$) whose result is consistent with Nickel's result.

With  increasing the magnetic field the DCDW phase  ($\xi=0,q\neq 0$), the HCC phase ($\xi\neq 0,q\neq 0$) 
and the chiral-restored phase  ($\xi=1$) appear.
In the low $\mu$ region, $q$ is small compared to other order parameters, so we call the DCDW phase in the low $\mu$ region "weak DCDW phase."
We call the DCDW phase in the high $\mu$ region "strong DCDW phase," which is similar to the usual DCDW phase with the wavevector $q$ being sufficiently large.
In other words we may say that the weak DCDW phase is driven by the topological effect due to spectral asymmetry, while the strong DCDW phase by the nesting effect.
We can see the second order phase transition between the weak DCDW phase to the HCC phase, where the order parameters are  continuously changed.
At the phase transition between  the HCC phase to  the strong DCDW phase, the order parameters  change discontinuously.  
Thus this phase transition is of the first order.

As we have already seen in the RKC phase, the order parameters should exhibit the de Haas-van Alphen effect \cite{deHaas1936,Ebert:1999ht} as a function of  the magnetic field.
In our results, the corresponding effect cannot be seen, since we have discarded the contribution of the higher Landau levels in our approximation.
If  full order contributions of the magnetic field is taken into account, the de Haas-van Alphen effect can appear.
Anyway, the oscillation of the order parameter should be very small at the weak $B$ region , where one may expect the HCC phase.
On the other hand, only the DCDW phase appears in the high $B$ region, and the phase diagram becomes the same with Frolov's results \cite{Frolov:2010wn}.

\section{ Summary and Concluding remarks\label{sec:concl}}
In this paper, we have studied chiral phase transition in the external magnetic field $\bf B$, 
taking account of a new type of the inhomogeneous condensate called hybrid chiral condensate (HCC).
HCC is then a self-consistent solution within the NJL model under the mean-field approximation, and exhibits both features of DCDW and RKC.
We have seen that the quark energy spectrum becomes asymmetric about zero due to the phase degree of freedom of DCDW, and there appear the mid-gap states 
due to the solitonic  property of RKC. 
Generally speaking, spectral asymmetry  plays very important roles for appearance of inhomogeneous phase in the magnetic field \cite{Tatsumi:2014wka}.
In  some case,  the contribution of spectral asymmetry is equivalent to  manifestation of chiral anomaly.
We have explicitly evaluated the Atiyah-Patodi-Singer $\eta$-invariant for the case of HCC.
The energy spectrum has a gap and $\eta_H$ is given by the sum of the one given by the states above and below the gap and the one given by the mid-gap states;
the former does not depend on the modulus parameter in HCC to give the same form as in the DCDW case,
and the latter is related to the number of nodes of HCC. 

 We have studied the phase diagram of the inhomogeneous chiral phase in the $\mu-B$ plane at $T=0$ for two cases. 
First we have explored the pure RKC phase and found that the phase diagram is little affected by the magnetic field, 
while some oscillation due to the de Haas-van Alphen effect can be slightly observed.
Next, we have discussed the full phase diagram, taking into account HCC. 
Since spectral asymmetry implies that the phase modulation is always favored in the magnetic field, 
independent of $\mu$, 
the phase diagram consists of three regions besides the chiral-restored phase: the strong DCDW phase, the weak DCDW phase, and the HCC phase.
The strong DCDW phase resembles the pure DCDW phase, but the appearance of the weak DCDW phase is attributed to the anomalous quark number caused by spectral asymmetry; 
actually they disappear as the magnetic field is turned off. 
Note that pure RKC phase never appears once the magnetic field is turned on, and is replaced by the HCC phase.



In this paper, we have considered only the flavor symmetric case, $\mu_u=\mu_d$ for simplicity,
while $u$ and $d$ quarks  should have different number in a realistic situation due to different electric charge.
 Actually cold catalyzed matter develops inside neutron stars, where charge neutrality and chemical equilibrium should be established.
Thus nonzero isospin chemical potential ($\mu_{\rm I}\equiv\mu_u-\mu_d\neq 0$) is very important in the magnetic field.
In Refs. \cite{Ebert:2014woa,Abuki:2013vwa,Abuki:2013pla}, 
they have been studied the phase diagram taking into account isospin chemical potential in the absence of the magnetic field.

We  have also considered  cold quark matter ($T=0$), while it may be interesting to study how thermal effect modifies our findings  .
Actually it has been recently discussed that the external magnetic field suppresses chiral condensates at finite temperature
 \cite{Bali:2011qj,Fukushima:2012kc,Mueller:2015fka}.
This subject is to be discussed elsewhere.

We have discussed the phase diagram of the inhomogeneous chiral phase in the chiral limit $m_c=0$,
while it is known that the current quark mass $m_c$ suppresses the inhomogeneous phase \cite{Nickel:2009wj,Karasawa:2013zsa,Maedan:2009yi,Abuki:2011pf}.
For the case with the magnetic field it has been supposed that the effect of current quark mass defeat the effect of spectral asymmetry and that chiral condensate is homogeneous at low $\mu$ and low $B$ region \cite{Tatsumi:2014wka}.

Finally it should be worth mentioning that HCC may have some implications in the context of the FFLO state of superconductivity \cite{Fulde:1964zz,larkin:1964zz}.
Very recently an evidence of the LO state has been reported, and the Andreev bound states are emphasized as a hallmark \cite{Mayaffre:2014}. 
Usually this subject has been separately discussed for the FF state or the LO state.
Since it has been shown that there is a duality relation between superconductivity and spontaneous breaking of chiral symmetry in 1+1 dimensions \cite{Thies:2003zr},
the FFLO state with one dimensional modulation may be similarly treated to our subject;
the Andreev bound states then correspond to the mid-gap states.
Note that HCC satisfies the BdG equation and  can give the pairing function connecting the LO and FF states smoothly.
The coexistence of the FF and LO states in the quasi-one dimensional system has been  discussed as an appearance of time crystal phase in which time translation symmetry is spontaneously broken \cite{Yoshii:2014fwa}.

\section*{Acknowledgement}
We thank N. Yamanaka and R. Yoshiike and T.-G. Lee for useful discussions. This work is partially supported by Grants-in-Aid for Scientific Research on Innovative Areas 
through No. 24105008 provided by MEXT.

 \appendix
\section{Regularization of $ \Omega_{\rm vac} $\label{sec:reg}}

Since $\Omega _{\rm vac}$ is divergent, we apply the  proper time regularization for $ \Omega_{\rm vac} $.
At $\mu=T=0$, thermodynamic potential is written in
\begin{eqnarray*}
 \Omega_{\rm vac} &=& -\frac{1}{2}N_c \sum_{f}\frac{|e_fB|}{2\pi}\sum_{n,\zeta}\int dp_4 \int d\lambda \rho_\zeta(\lambda){\rm ln}\left({\cal E}^2+p_4^2 \right)
\end{eqnarray*}
For ${\rm Re}A>0$ the equation, 
\begin{eqnarray}
\frac{1}{A^x} &=& \frac{1}{(x-1)!}\int^{\infty}_{0} d\tau ~\tau^{x-1}e^{-\tau A},
\end{eqnarray}
holds.

Thermodynamic potential then becomes
\begin{eqnarray*}
\Omega _{\rm vac} &=& N_c \sum_{f} \sum_{n,\zeta}\frac{|e_fB|}{8\pi^{3/2}}
\int^{\infty}_{-\infty}d\lambda \rho_\zeta(\lambda) 
\int_{0}^{\infty}\frac{d\tau}{\tau^{3/2}}{\rm exp}\left[-\tau\left( \lambda^2+2|e_fB|n\right)\right]\\
&\to& N_c \sum_{f} \sum_{n,\zeta}\frac{|e_fB|}{8\pi^{3/2}}\int^{\infty}_{-\infty}d\lambda \rho_\zeta(\lambda) \int_{1/\Lambda^2}^{\infty}\frac{d\tau}{\tau^{3/2}}{\rm exp}\left[-\tau\left( \lambda^2+2|e_fB|n\right)\right]\\
&=& N_c \sum_{f,\zeta} \frac{|e_fB|}{16\pi^{3/2}}\int_{1/\Lambda^2}^{\infty}\frac{d\tau}{\tau^{3/2}}{\rm coth}(\tau|e_fB|)
\int^{\infty}_{-\infty}d\lambda \rho_\zeta(\lambda){\rm exp}\left(-\tau\lambda^2\right)\\
\end{eqnarray*}

\section{Some remarks on spectral asymmetry\label{sec:spectral2}}
Spectral asymmetry is closely related to axial anomaly in the specific case \cite{Tatsumi:2014wka}.
In the effective theory of mesons, anomalous contribution  coming from the Wess-Zumino-Witten (WZW) term is given by \cite{Son:2007ny}
\begin{eqnarray}
S_{WZW} &=& \frac{e}{4\pi^2 f_{\pi}}\int d^4x \mu {\bf B \cdot \nabla}\pi^0\\
\nonumber f_\pi&=&\sigma^2+(\pi^0)^2
\end{eqnarray}
 in the presence of magnetic field and chemical potential. Our DCDW configuration  may correspond to
\begin{eqnarray}
\nonumber \sigma + i \pi^0  = f_\pi e^{iqx},
\end{eqnarray}
 in this context.
For this configuration,  the WZW term reads
\begin{eqnarray}
S_{WZW} &=& \frac{e\mu}{4\pi^2 }\int d^4x {\bf B \cdot q}
\end{eqnarray}
 Accordingly the anomalous number density is given by  
\begin{eqnarray}
n &=& \frac{e}{4\pi^2}{\bf B \cdot q},
\end{eqnarray}
which is the same form as Eq.~(\ref{eta-DCDW}).
This implies that ${\bf B \parallel q}$ is always favorable.

\section{Spectral asymmetry for HCC \label{etainv}}

Consider the LLL. The $\eta$ invariant is then given by 
\begin{equation}
\eta_H = VN_c \sum_{f}\frac{|e_fB|}{2\pi}\lim_{s \to +0} \int_{-\infty}^{\infty}d\lambda \rho_+(\lambda){\rm sign}(\lambda)\left|\lambda\right|^{-s},
\label{aps}
\end{equation}
with  the density of states Eq.~(\ref{eq:density}).

For the case, $q/2<m$, for simplicity, the integral (\ref{aps}) is divided into two parts:
\begin{eqnarray}
\nonumber \left(\int_{-\infty}^{E_1} +\int_{E_2-q/2}^{E_3-q/2}+\int_{E_4}^\infty\right)d\lambda \rho_+(\lambda){\rm sign}(\lambda)\left|\lambda\right|^{-s}
&=& m\int_1^\infty dx \frac{1}{\pi}\frac{x^2 + c}{\sqrt{(x^2-1)(x^2-\nu')}}\left[\left(mx+\frac{q}{2}\right)^{-s}-\left(mx-\frac{q}{2}\right)^{-s}\right]\\
&&+N_{\rm midgap}.
\label{int1}
\end{eqnarray}
The second integral in Eq.~({\ref{int1}})  is the contribution of the mid-gap states, 
 \begin{eqnarray}
\nonumber N_{\rm midgap} &=&\int^{E_3}_{E_2} d\lambda \rho_+(\lambda){\rm sign}(\lambda).
\end{eqnarray}
For $m\nu'<q/2$, $N_{\rm midgap}$ is equal to the number of nodes of HCC: $N_{\rm nodes}=m/[(1+\sqrt{\nu}){\bf K}(\nu)]$.
Using the incomplete elliptic integrals, $N_{\rm midgap}$ is written in $\frac{2m}{\pi}{\rm Re}\left({\bf F}(q/2;\nu^\prime) +c{\bf F}(q/2;\nu^\prime)-{\bf E}(q/2;\nu^\prime) \right)$
for $m\nu'>q/2$.

In the following, we consider the first term. Expanding it with respect to $q$, we have
\begin{equation}
-2m\int_m^\infty dx \frac{1}{\pi}\frac{x^2 + c}{\sqrt{(x^2-1)(x^2-\nu')}}\left[s(mx)^{-(s+1)}\frac{q}{2}+\frac{1}{6}s(s+1)(s+2)(mx)^{-(s+3)}\left(\frac{q}{2}\right)^2+O(q^5)\right].
\end{equation}
Since other terms become zero as $s\rightarrow 0$, we, hereafter, evaluate only the first term,
\begin{eqnarray}
-sqm^{-s}\int_1^\infty dx\frac{1}{\pi}\frac{x^2 + c}{\sqrt{(x^2-1)(x^2-\nu')}}.
\end{eqnarray}
Transforming the integration variable $x$ by $t=x^{-2}$, we have 
\begin{eqnarray}
&&-sqm^{-s}\frac{1}{2\pi}\int_0^1 dt t^{s/2-1}(1-t)^{-1/2}(1-\nu' t)^{-1/2}(1+ct)\nonumber\\
&=&-sqm^{-s}\frac{1}{2\pi}\frac{\Gamma(s/2)\Gamma(1/2)}{\Gamma((1+s)/2)}F(1/2,s/2,(1+s)/2;\nu')+({\rm regular~terms ~in}~s),
\end{eqnarray}
in terms of the Gauss hypergeometric function $F$. Using the relation, $\Gamma(s/2)=2/s\Gamma(s/2+1)$, and taking the limit $s\rightarrow 0$,
we have 
\begin{equation}
-\frac{q}{\pi}F(1/2,0,1/2;\nu')=-\frac{q}{\pi}
\end{equation}
Thus the $\eta$ invariant can be given as
\begin{equation}
\eta_H=-VN_c \sum_{f}\frac{|e_fB|}{2\pi}\left(\frac{q}{\pi}-N_{\rm midgap}\right),
\label{eta-hcc}
\end{equation}
for the case, $q/2<m$.  It is easy to evaluate the $\eta$-invariant for $q/2>m$.

\begin{figure}[htb]
\begin{center}
\subfigure[DCDW ($q=$const.)]{\includegraphics[width=5cm,angle=0]{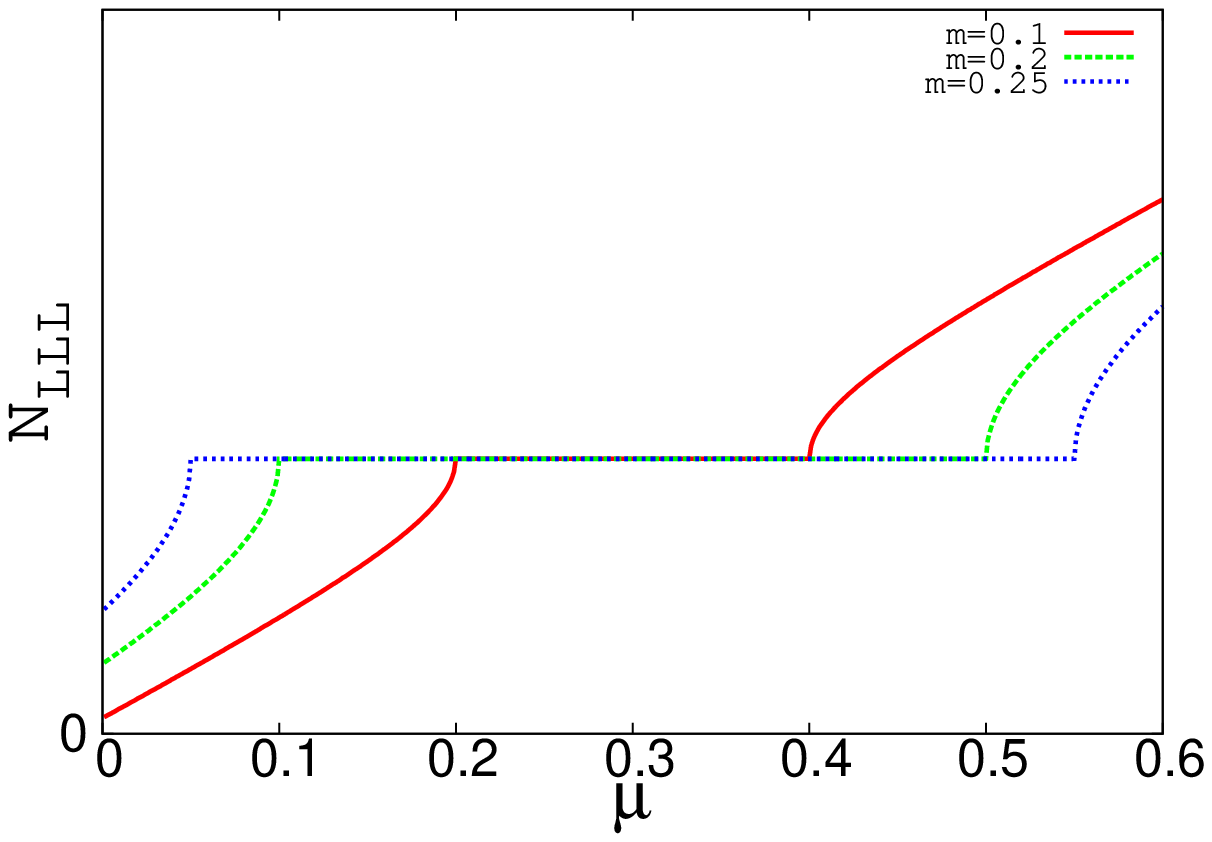}}
\subfigure[HCC (the number of nodes is fixed)]{\includegraphics[width=5cm,angle=0]{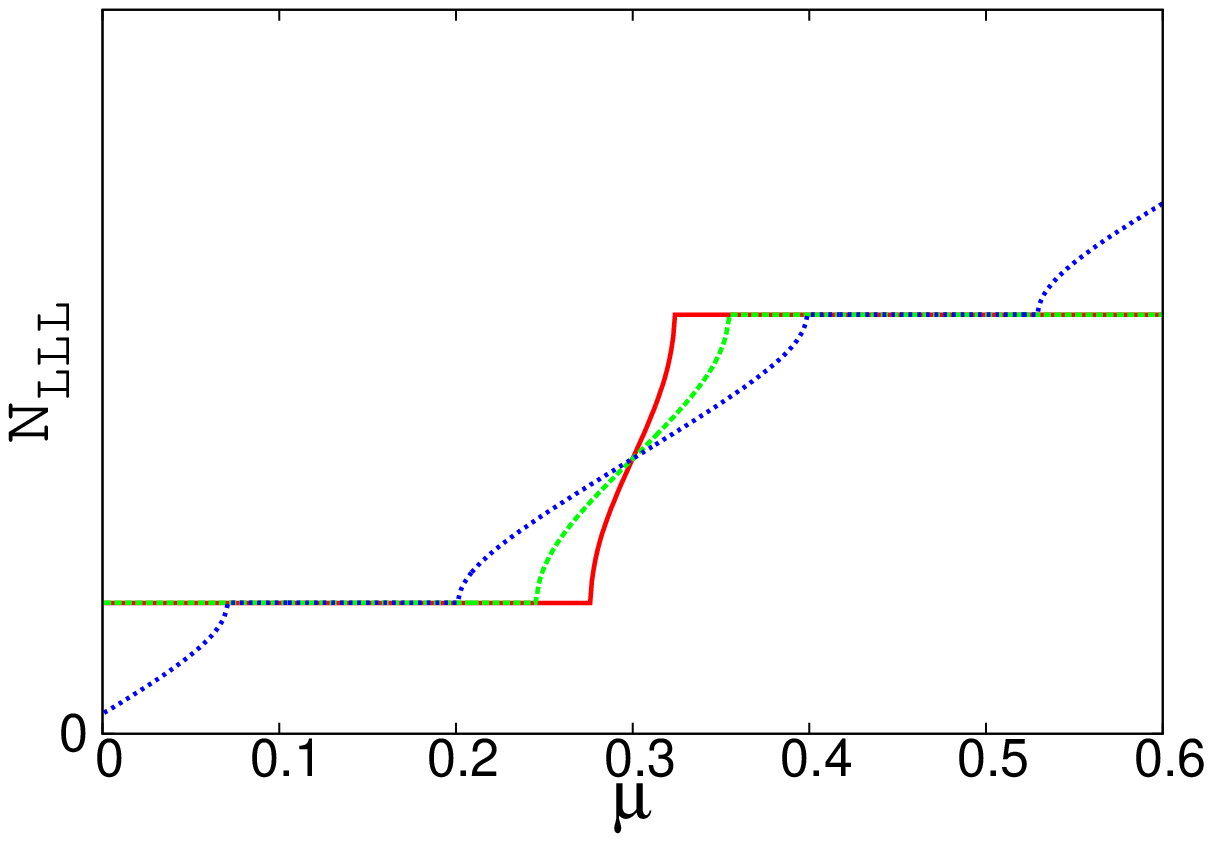}}
   \end{center}
\caption{The behavior of $N_{\rm LLL}/V$ as a function of $\mu$ for different conditions.
}
\label{fig:rholll}
\end{figure}

Figure \ref{fig:rholll} shows the behavior of the number of the occupied states in LLL, $N_{\rm LLL}$, which consists of the normal baryon number density and $\eta$-invariant.
For DCDW, the value of $N_{\rm LLL}$  in the plateau is independent of $m$.
For HCC, the value of $N_{\rm LLL}$ in the plateau depends on $m$ and $\nu$.
When the number of nodes is fixed, the value of $N_{\rm LLL}$ in the plateau depends on only $q$.

\section{Expansion of thermodynamic potential with respect to the magnetic field \label{sec:expansion}}
Before summation over the Matsubara frequencies, $\omega_l = (2n+1)\pi T$ , the thermodynamic potential can be written as
\begin{eqnarray}
\nonumber \Omega_{1p} &=&-\frac12TN_c\sum_f\frac{|e_fB|}{2\pi}\sum_{l,n,\zeta}\int d\lambda \rho_\zeta(\lambda){\rm ln}\left[ \omega_l^2 + (\lambda\sqrt{1+\frac{2|e_fB|n}{\lambda^2}} - \mu )^2 \right] \\
 \nonumber &=&-\frac12TN_c\sum_f\frac{|e_fB|}{2\pi}\sum_{l,n,\zeta}\int d\lambda \rho_\zeta(\lambda)(1-\frac{1}{2}\delta_{n,0}){\rm ln}\left[ (\omega_l -i\mu)^2 + \lambda^2+2|e_fB|n \right] \\
 \nonumber & &-\frac14TN_c\sum_f\frac{|e_fB|}{2\pi}\sum_{l}\int  d\lambda \left[ \rho_+(\lambda) - \rho_-(\lambda)\right]{\rm ln}\left[ \omega_l^2 + \left( \lambda -\mu \right)^2 \right] \\
 &=&\Omega_{\rm even}+\Omega_{\rm odd}.
\label{eq:tpmatsubara}
\end{eqnarray}
Only LLL contributes to $\Omega_{\rm odd}$ that is the first order term of $q$.
\begin{eqnarray}
\nonumber \Omega_{\rm even} &=&
-\frac12TN_c\sum_f\frac{|e_fB|}{2\pi}\sum_{l,n,\zeta}\int d\lambda\rho_\zeta(\lambda)\left(1-\frac{1}{2}\delta_{n,0}\right){\rm ln}\left[ (\omega_l -i\mu)^2 + \lambda^2+2|e_fB|n \right] \\
 &=&-\frac12TN_c\sum_f\frac{|e_fB|}{2\pi}\sum_{l,n,\zeta}\int d\lambda\rho_\zeta(\lambda)\left(1-\frac{1}{2}\delta_{n,0}\right)\int^\infty_0 \frac{d\tau}{\tau} e^{-\tau \left[ (\omega_l -i\mu)^2 + \lambda^2+2|e_fB|n \right]}\\
 &=&-\frac12TN_c\sum_f\frac{|e_fB|}{2\pi}\sum_{l,\zeta}\int d\lambda\rho_\zeta(\lambda)\int^\infty_0 \frac{d\tau}{\tau} {\rm coth}\left(-\tau |e_fB| \right)e^{-\tau \left[ (\omega_l -i\mu)^2 + \lambda^2\right]}.
\end{eqnarray}
Since $x {\rm coth}(x)$ is the even function of $x$, $\Omega_{\rm even}$ contains only even order terms of $B$.

\bibliographystyle{apsrev}
\bibliography{./ref}

\begin{thebibliography}{49}
\expandafter\ifx\csname natexlab\endcsname\relax\def\natexlab#1{#1}\fi
\expandafter\ifx\csname bibnamefont\endcsname\relax
  \def\bibnamefont#1{#1}\fi
\expandafter\ifx\csname bibfnamefont\endcsname\relax
  \def\bibfnamefont#1{#1}\fi
\expandafter\ifx\csname citenamefont\endcsname\relax
  \def\citenamefont#1{#1}\fi
\expandafter\ifx\csname url\endcsname\relax
  \def\url#1{\texttt{#1}}\fi
\expandafter\ifx\csname urlprefix\endcsname\relax\def\urlprefix{URL }\fi
\providecommand{\bibinfo}[2]{#2}
\providecommand{\eprint}[2][]{\url{#2}}

\bibitem[{\citenamefont{Nakano and Tatsumi}(2005)}]{Nakano:2004cd}
\bibinfo{author}{\bibfnamefont{E.}~\bibnamefont{Nakano}} \bibnamefont{and}
  \bibinfo{author}{\bibfnamefont{T.}~\bibnamefont{Tatsumi}},
  \bibinfo{journal}{Phys.Rev.} \textbf{\bibinfo{volume}{D71}},
  \bibinfo{pages}{114006} (\bibinfo{year}{2005}), \eprint{hep-ph/0411350}.

\bibitem[{\citenamefont{Tatsumi et~al.}(2015)\citenamefont{Tatsumi, Nishiyama,
  and Karasawa}}]{Tatsumi:2014wka}
\bibinfo{author}{\bibfnamefont{T.}~\bibnamefont{Tatsumi}},
  \bibinfo{author}{\bibfnamefont{K.}~\bibnamefont{Nishiyama}},
  \bibnamefont{and} \bibinfo{author}{\bibfnamefont{S.}~\bibnamefont{Karasawa}},
  \bibinfo{journal}{Phys.Lett.} \textbf{\bibinfo{volume}{B743}},
  \bibinfo{pages}{66} (\bibinfo{year}{2015}), \eprint{1405.2155}.

\bibitem[{\citenamefont{Schon and Thies}(2000)}]{Schon:2000qy}
\bibinfo{author}{\bibfnamefont{V.}~\bibnamefont{Schon}} \bibnamefont{and}
  \bibinfo{author}{\bibfnamefont{M.}~\bibnamefont{Thies}}
  (\bibinfo{year}{2000}), \eprint{hep-th/0008175}.

\bibitem[{\citenamefont{Thies and Urlichs}(2003)}]{Thies:2003kk}
\bibinfo{author}{\bibfnamefont{M.}~\bibnamefont{Thies}} \bibnamefont{and}
  \bibinfo{author}{\bibfnamefont{K.}~\bibnamefont{Urlichs}},
  \bibinfo{journal}{Phys.Rev.} \textbf{\bibinfo{volume}{D67}},
  \bibinfo{pages}{125015} (\bibinfo{year}{2003}), \eprint{hep-th/0302092}.

\bibitem[{\citenamefont{Nickel}(2009{\natexlab{a}})}]{Nickel:2009wj}
\bibinfo{author}{\bibfnamefont{D.}~\bibnamefont{Nickel}},
  \bibinfo{journal}{Phys.Rev.} \textbf{\bibinfo{volume}{D80}},
  \bibinfo{pages}{074025} (\bibinfo{year}{2009}{\natexlab{a}}),
  \eprint{0906.5295}.

\bibitem[{\citenamefont{Nickel}(2009{\natexlab{b}})}]{Nickel:2009ke}
\bibinfo{author}{\bibfnamefont{D.}~\bibnamefont{Nickel}},
  \bibinfo{journal}{Phys.Rev.Lett.} \textbf{\bibinfo{volume}{103}},
  \bibinfo{pages}{072301} (\bibinfo{year}{2009}{\natexlab{b}}),
  \eprint{0902.1778}.

\bibitem[{\citenamefont{Basar and Dunne}(2008)}]{Basar:2008ki}
\bibinfo{author}{\bibfnamefont{G.}~\bibnamefont{Basar}} \bibnamefont{and}
  \bibinfo{author}{\bibfnamefont{G.~V.} \bibnamefont{Dunne}},
  \bibinfo{journal}{Phys.Rev.} \textbf{\bibinfo{volume}{D78}},
  \bibinfo{pages}{065022} (\bibinfo{year}{2008}), \eprint{0806.2659}.

\bibitem[{\citenamefont{Basar et~al.}(2009)\citenamefont{Basar, Dunne, and
  Thies}}]{Basar:2009fg}
\bibinfo{author}{\bibfnamefont{G.}~\bibnamefont{Basar}},
  \bibinfo{author}{\bibfnamefont{G.~V.} \bibnamefont{Dunne}}, \bibnamefont{and}
  \bibinfo{author}{\bibfnamefont{M.}~\bibnamefont{Thies}},
  \bibinfo{journal}{Phys.Rev.} \textbf{\bibinfo{volume}{D79}},
  \bibinfo{pages}{105012} (\bibinfo{year}{2009}), \eprint{0903.1868}.

\bibitem[{\citenamefont{Kojo et~al.}(2010)\citenamefont{Kojo, Hidaka, McLerran,
  and Pisarski}}]{Kojo:2009ha}
\bibinfo{author}{\bibfnamefont{T.}~\bibnamefont{Kojo}},
  \bibinfo{author}{\bibfnamefont{Y.}~\bibnamefont{Hidaka}},
  \bibinfo{author}{\bibfnamefont{L.}~\bibnamefont{McLerran}}, \bibnamefont{and}
  \bibinfo{author}{\bibfnamefont{R.~D.} \bibnamefont{Pisarski}},
  \bibinfo{journal}{Nucl.Phys.} \textbf{\bibinfo{volume}{A843}},
  \bibinfo{pages}{37} (\bibinfo{year}{2010}), \eprint{0912.3800}.

\bibitem[{\citenamefont{Kojo et~al.}(2012)\citenamefont{Kojo, Hidaka,
  Fukushima, McLerran, and Pisarski}}]{Kojo:2011cn}
\bibinfo{author}{\bibfnamefont{T.}~\bibnamefont{Kojo}},
  \bibinfo{author}{\bibfnamefont{Y.}~\bibnamefont{Hidaka}},
  \bibinfo{author}{\bibfnamefont{K.}~\bibnamefont{Fukushima}},
  \bibinfo{author}{\bibfnamefont{L.~D.} \bibnamefont{McLerran}},
  \bibnamefont{and} \bibinfo{author}{\bibfnamefont{R.~D.}
  \bibnamefont{Pisarski}}, \bibinfo{journal}{Nucl.Phys.}
  \textbf{\bibinfo{volume}{A875}}, \bibinfo{pages}{94} (\bibinfo{year}{2012}),
  \eprint{1107.2124}.

\bibitem[{\citenamefont{M{\"u}ller et~al.}(2013)\citenamefont{M{\"u}ller,
  Buballa, and Wambach}}]{Muller:2013tya}
\bibinfo{author}{\bibfnamefont{D.}~\bibnamefont{M{\"u}ller}},
  \bibinfo{author}{\bibfnamefont{M.}~\bibnamefont{Buballa}}, \bibnamefont{and}
  \bibinfo{author}{\bibfnamefont{J.}~\bibnamefont{Wambach}},
  \bibinfo{journal}{Phys.Lett.} \textbf{\bibinfo{volume}{B727}},
  \bibinfo{pages}{240} (\bibinfo{year}{2013}), \eprint{1308.4303}.

\bibitem[{\citenamefont{Buballa and Carignano}(2015)}]{Buballa:2014tba}
\bibinfo{author}{\bibfnamefont{M.}~\bibnamefont{Buballa}} \bibnamefont{and}
  \bibinfo{author}{\bibfnamefont{S.}~\bibnamefont{Carignano}},
  \bibinfo{journal}{Prog.Part.Nucl.Phys.} \textbf{\bibinfo{volume}{81}},
  \bibinfo{pages}{39} (\bibinfo{year}{2015}), \eprint{1406.1367}.

\bibitem[{\citenamefont{Carignano et~al.}(2010)\citenamefont{Carignano, Nickel,
  and Buballa}}]{Carignano:2010ac}
\bibinfo{author}{\bibfnamefont{S.}~\bibnamefont{Carignano}},
  \bibinfo{author}{\bibfnamefont{D.}~\bibnamefont{Nickel}}, \bibnamefont{and}
  \bibinfo{author}{\bibfnamefont{M.}~\bibnamefont{Buballa}},
  \bibinfo{journal}{Phys.Rev.} \textbf{\bibinfo{volume}{D82}},
  \bibinfo{pages}{054009} (\bibinfo{year}{2010}), \eprint{1007.1397}.

\bibitem[{\citenamefont{Carignano et~al.}(2014)\citenamefont{Carignano,
  Buballa, and Schaefer}}]{Carignano:2014jla}
\bibinfo{author}{\bibfnamefont{S.}~\bibnamefont{Carignano}},
  \bibinfo{author}{\bibfnamefont{M.}~\bibnamefont{Buballa}}, \bibnamefont{and}
  \bibinfo{author}{\bibfnamefont{B.-J.} \bibnamefont{Schaefer}},
  \bibinfo{journal}{Phys.Rev.} \textbf{\bibinfo{volume}{D90}},
  \bibinfo{pages}{014033} (\bibinfo{year}{2014}), \eprint{1404.0057}.

\bibitem[{\citenamefont{Carignano and Buballa}(2012)}]{Carignano:2012sx}
\bibinfo{author}{\bibfnamefont{S.}~\bibnamefont{Carignano}} \bibnamefont{and}
  \bibinfo{author}{\bibfnamefont{M.}~\bibnamefont{Buballa}},
  \bibinfo{journal}{Phys.Rev.} \textbf{\bibinfo{volume}{D86}},
  \bibinfo{pages}{074018} (\bibinfo{year}{2012}), \eprint{1203.5343}.

\bibitem[{\citenamefont{Karasawa and Tatsumi}(2013)}]{Karasawa:2013zsa}
\bibinfo{author}{\bibfnamefont{S.}~\bibnamefont{Karasawa}} \bibnamefont{and}
  \bibinfo{author}{\bibfnamefont{T.}~\bibnamefont{Tatsumi}}
  (\bibinfo{year}{2013}), \eprint{1307.6448}.

\bibitem[{\citenamefont{Maedan}(2010)}]{Maedan:2009yi}
\bibinfo{author}{\bibfnamefont{S.}~\bibnamefont{Maedan}},
  \bibinfo{journal}{Prog.Theor.Phys.} \textbf{\bibinfo{volume}{123}},
  \bibinfo{pages}{285} (\bibinfo{year}{2010}), \eprint{0908.0594}.

\bibitem[{\citenamefont{Abuki}(2013)}]{Abuki:2013vwa}
\bibinfo{author}{\bibfnamefont{H.}~\bibnamefont{Abuki}},
  \bibinfo{journal}{Phys.Rev.} \textbf{\bibinfo{volume}{D87}},
  \bibinfo{pages}{094006} (\bibinfo{year}{2013}), \eprint{1304.1904}.

\bibitem[{\citenamefont{Abuki}(2014)}]{Abuki:2013pla}
\bibinfo{author}{\bibfnamefont{H.}~\bibnamefont{Abuki}},
  \bibinfo{journal}{Phys.Lett.} \textbf{\bibinfo{volume}{B728}},
  \bibinfo{pages}{427} (\bibinfo{year}{2014}), \eprint{1307.8173}.

\bibitem[{\citenamefont{Abuki et~al.}(2012)\citenamefont{Abuki, Ishibashi, and
  Suzuki}}]{Abuki:2011pf}
\bibinfo{author}{\bibfnamefont{H.}~\bibnamefont{Abuki}},
  \bibinfo{author}{\bibfnamefont{D.}~\bibnamefont{Ishibashi}},
  \bibnamefont{and} \bibinfo{author}{\bibfnamefont{K.}~\bibnamefont{Suzuki}},
  \bibinfo{journal}{Phys.Rev.} \textbf{\bibinfo{volume}{D85}},
  \bibinfo{pages}{074002} (\bibinfo{year}{2012}), \eprint{1109.1615}.

\bibitem[{\citenamefont{Frolov et~al.}(2010)\citenamefont{Frolov, Zhukovsky,
  and Klimenko}}]{Frolov:2010wn}
\bibinfo{author}{\bibfnamefont{I.}~\bibnamefont{Frolov}},
  \bibinfo{author}{\bibfnamefont{V.~C.} \bibnamefont{Zhukovsky}},
  \bibnamefont{and} \bibinfo{author}{\bibfnamefont{K.}~\bibnamefont{Klimenko}},
  \bibinfo{journal}{Phys.Rev.} \textbf{\bibinfo{volume}{D82}},
  \bibinfo{pages}{076002} (\bibinfo{year}{2010}), \eprint{1007.2984}.

\bibitem[{\citenamefont{Gubina et~al.}(2012)\citenamefont{Gubina, Klimenko,
  Kurbanov, and Zhukovsky}}]{Gubina:2012wp}
\bibinfo{author}{\bibfnamefont{N.}~\bibnamefont{Gubina}},
  \bibinfo{author}{\bibfnamefont{K.}~\bibnamefont{Klimenko}},
  \bibinfo{author}{\bibfnamefont{S.}~\bibnamefont{Kurbanov}}, \bibnamefont{and}
  \bibinfo{author}{\bibfnamefont{V.~C.} \bibnamefont{Zhukovsky}},
  \bibinfo{journal}{Phys.Rev.} \textbf{\bibinfo{volume}{D86}},
  \bibinfo{pages}{085011} (\bibinfo{year}{2012}), \eprint{1206.2519}.

\bibitem[{\citenamefont{Ebert et~al.}(2014)\citenamefont{Ebert, Khunjua,
  Klimenko, and Zhukovsky}}]{Ebert:2014woa}
\bibinfo{author}{\bibfnamefont{D.}~\bibnamefont{Ebert}},
  \bibinfo{author}{\bibfnamefont{T.}~\bibnamefont{Khunjua}},
  \bibinfo{author}{\bibfnamefont{K.}~\bibnamefont{Klimenko}}, \bibnamefont{and}
  \bibinfo{author}{\bibfnamefont{V.~C.} \bibnamefont{Zhukovsky}},
  \bibinfo{journal}{Phys.Rev.} \textbf{\bibinfo{volume}{D90}},
  \bibinfo{pages}{045021} (\bibinfo{year}{2014}), \eprint{1405.3789}.

\bibitem[{\citenamefont{Overhauser}(1962)}]{Overhauser:1962zz}
\bibinfo{author}{\bibfnamefont{A.}~\bibnamefont{Overhauser}},
  \bibinfo{journal}{Phys.Rev.} \textbf{\bibinfo{volume}{128}},
  \bibinfo{pages}{1437} (\bibinfo{year}{1962}).

\bibitem[{\citenamefont{Pierls}(1955)}]{Pierls:1955}
\bibinfo{author}{\bibfnamefont{P.~E.} \bibnamefont{Pierls}},
  \emph{\bibinfo{title}{Quantum Theory of Solids}}
  (\bibinfo{publisher}{Clarendon Press}, \bibinfo{year}{1955}).

\bibitem[{\citenamefont{Fulde and Ferrell}(1964)}]{Fulde:1964zz}
\bibinfo{author}{\bibfnamefont{P.}~\bibnamefont{Fulde}} \bibnamefont{and}
  \bibinfo{author}{\bibfnamefont{R.~A.} \bibnamefont{Ferrell}},
  \bibinfo{journal}{Phys.Rev.} \textbf{\bibinfo{volume}{135}},
  \bibinfo{pages}{A550} (\bibinfo{year}{1964}).

\bibitem[{\citenamefont{Larkin and Ovchinnikov}(1964)}]{larkin:1964zz}
\bibinfo{author}{\bibfnamefont{A.}~\bibnamefont{Larkin}} \bibnamefont{and}
  \bibinfo{author}{\bibfnamefont{Y.}~\bibnamefont{Ovchinnikov}},
  \bibinfo{journal}{Zh.Eksp.Teor.Fiz.} \textbf{\bibinfo{volume}{47}},
  \bibinfo{pages}{1136} (\bibinfo{year}{1964}).

\bibitem[{\citenamefont{Skokov et~al.}(2009)\citenamefont{Skokov, Illarionov,
  and Toneev}}]{Skokov:2009qp}
\bibinfo{author}{\bibfnamefont{V.}~\bibnamefont{Skokov}},
  \bibinfo{author}{\bibfnamefont{A.~Y.} \bibnamefont{Illarionov}},
  \bibnamefont{and} \bibinfo{author}{\bibfnamefont{V.}~\bibnamefont{Toneev}},
  \bibinfo{journal}{Int.J.Mod.Phys.} \textbf{\bibinfo{volume}{A24}},
  \bibinfo{pages}{5925} (\bibinfo{year}{2009}), \eprint{0907.1396}.

\bibitem[{\citenamefont{Fukushima et~al.}(2008)\citenamefont{Fukushima,
  Kharzeev, and Warringa}}]{Fukushima:2008xe}
\bibinfo{author}{\bibfnamefont{K.}~\bibnamefont{Fukushima}},
  \bibinfo{author}{\bibfnamefont{D.~E.} \bibnamefont{Kharzeev}},
  \bibnamefont{and} \bibinfo{author}{\bibfnamefont{H.~J.}
  \bibnamefont{Warringa}}, \bibinfo{journal}{Phys.Rev.}
  \textbf{\bibinfo{volume}{D78}}, \bibinfo{pages}{074033}
  (\bibinfo{year}{2008}), \eprint{0808.3382}.

\bibitem[{\citenamefont{Gusynin et~al.}(1996)\citenamefont{Gusynin, Miransky,
  and Shovkovy}}]{Gusynin:1995nb}
\bibinfo{author}{\bibfnamefont{V.}~\bibnamefont{Gusynin}},
  \bibinfo{author}{\bibfnamefont{V.}~\bibnamefont{Miransky}}, \bibnamefont{and}
  \bibinfo{author}{\bibfnamefont{I.}~\bibnamefont{Shovkovy}},
  \bibinfo{journal}{Nucl.Phys.} \textbf{\bibinfo{volume}{B462}},
  \bibinfo{pages}{249} (\bibinfo{year}{1996}), \eprint{hep-ph/9509320}.

\bibitem[{\citenamefont{Suganuma and Tatsumi}(1991)}]{Suganuma:1990nn}
\bibinfo{author}{\bibfnamefont{H.}~\bibnamefont{Suganuma}} \bibnamefont{and}
  \bibinfo{author}{\bibfnamefont{T.}~\bibnamefont{Tatsumi}},
  \bibinfo{journal}{Annals Phys.} \textbf{\bibinfo{volume}{208}},
  \bibinfo{pages}{470} (\bibinfo{year}{1991}).

\bibitem[{\citenamefont{Klevansky and Lemmer}(1989)}]{Klevansky:1989vi}
\bibinfo{author}{\bibfnamefont{S.}~\bibnamefont{Klevansky}} \bibnamefont{and}
  \bibinfo{author}{\bibfnamefont{R.~H.} \bibnamefont{Lemmer}},
  \bibinfo{journal}{Phys.Rev.} \textbf{\bibinfo{volume}{D39}},
  \bibinfo{pages}{3478} (\bibinfo{year}{1989}).

\bibitem[{\citenamefont{Bali et~al.}(2012)\citenamefont{Bali, Bruckmann,
  Endrodi, Fodor, Katz et~al.}}]{Bali:2011qj}
\bibinfo{author}{\bibfnamefont{G.}~\bibnamefont{Bali}},
  \bibinfo{author}{\bibfnamefont{F.}~\bibnamefont{Bruckmann}},
  \bibinfo{author}{\bibfnamefont{G.}~\bibnamefont{Endrodi}},
  \bibinfo{author}{\bibfnamefont{Z.}~\bibnamefont{Fodor}},
  \bibinfo{author}{\bibfnamefont{S.}~\bibnamefont{Katz}}, \bibnamefont{et~al.},
  \bibinfo{journal}{JHEP} \textbf{\bibinfo{volume}{1202}}, \bibinfo{pages}{044}
  (\bibinfo{year}{2012}), \eprint{1111.4956}.

\bibitem[{\citenamefont{Fukushima and Hidaka}(2013)}]{Fukushima:2012kc}
\bibinfo{author}{\bibfnamefont{K.}~\bibnamefont{Fukushima}} \bibnamefont{and}
  \bibinfo{author}{\bibfnamefont{Y.}~\bibnamefont{Hidaka}},
  \bibinfo{journal}{Phys.Rev.Lett.} \textbf{\bibinfo{volume}{110}},
  \bibinfo{pages}{031601} (\bibinfo{year}{2013}), \eprint{1209.1319}.

\bibitem[{\citenamefont{Braun et~al.}(2014)\citenamefont{Braun, Mian, and
  Rechenberger}}]{Braun:2014fua}
\bibinfo{author}{\bibfnamefont{J.}~\bibnamefont{Braun}},
  \bibinfo{author}{\bibfnamefont{W.~A.} \bibnamefont{Mian}}, \bibnamefont{and}
  \bibinfo{author}{\bibfnamefont{S.}~\bibnamefont{Rechenberger}}
  (\bibinfo{year}{2014}), \eprint{1412.6025}.

\bibitem[{\citenamefont{Mueller and Pawlowski}(2015)}]{Mueller:2015fka}
\bibinfo{author}{\bibfnamefont{N.}~\bibnamefont{Mueller}} \bibnamefont{and}
  \bibinfo{author}{\bibfnamefont{J.~M.} \bibnamefont{Pawlowski}}
  (\bibinfo{year}{2015}), \eprint{1502.08011}.

\bibitem[{\citenamefont{de~Haas and van Alphen}(1936)}]{deHaas1936}
\bibinfo{author}{\bibfnamefont{W.~J.} \bibnamefont{de~Haas}} \bibnamefont{and}
  \bibinfo{author}{\bibfnamefont{P.}~\bibnamefont{van Alphen}},
  \bibinfo{journal}{Proc. Am. Acad. Arts. Sci.} \textbf{\bibinfo{volume}{33}},
  \bibinfo{pages}{1106} (\bibinfo{year}{1936}).

\bibitem[{\citenamefont{Ebert et~al.}(2000)\citenamefont{Ebert, Klimenko,
  Vdovichenko, and Vshivtsev}}]{Ebert:1999ht}
\bibinfo{author}{\bibfnamefont{D.}~\bibnamefont{Ebert}},
  \bibinfo{author}{\bibfnamefont{K.}~\bibnamefont{Klimenko}},
  \bibinfo{author}{\bibfnamefont{M.}~\bibnamefont{Vdovichenko}},
  \bibnamefont{and}
  \bibinfo{author}{\bibfnamefont{A.}~\bibnamefont{Vshivtsev}},
  \bibinfo{journal}{Phys.Rev.} \textbf{\bibinfo{volume}{D61}},
  \bibinfo{pages}{025005} (\bibinfo{year}{2000}), \eprint{hep-ph/9905253}.

\bibitem[{\citenamefont{Nambu and Jona-Lasinio}(1961)}]{Nambu:1961tp}
\bibinfo{author}{\bibfnamefont{Y.}~\bibnamefont{Nambu}} \bibnamefont{and}
  \bibinfo{author}{\bibfnamefont{G.}~\bibnamefont{Jona-Lasinio}},
  \bibinfo{journal}{Phys.Rev.} \textbf{\bibinfo{volume}{122}},
  \bibinfo{pages}{345} (\bibinfo{year}{1961}).

\bibitem[{\citenamefont{Hatsuda and Kunihiro}(1994)}]{Hatsuda:1994pi}
\bibinfo{author}{\bibfnamefont{T.}~\bibnamefont{Hatsuda}} \bibnamefont{and}
  \bibinfo{author}{\bibfnamefont{T.}~\bibnamefont{Kunihiro}},
  \bibinfo{journal}{Phys.Rept.} \textbf{\bibinfo{volume}{247}},
  \bibinfo{pages}{221} (\bibinfo{year}{1994}), \eprint{hep-ph/9401310}.

\bibitem[{\citenamefont{Klevansky}(1992)}]{Klevansky:1992qe}
\bibinfo{author}{\bibfnamefont{S.}~\bibnamefont{Klevansky}},
  \bibinfo{journal}{Rev.Mod.Phys.} \textbf{\bibinfo{volume}{64}},
  \bibinfo{pages}{649} (\bibinfo{year}{1992}).

\bibitem[{\citenamefont{Sokolov and Ternov}(1986)}]{Sokolov1986}
\bibinfo{author}{\bibfnamefont{A.~A.} \bibnamefont{Sokolov}} \bibnamefont{and}
  \bibinfo{author}{\bibfnamefont{I.~M.} \bibnamefont{Ternov}},
  \emph{\bibinfo{title}{Radiation from Relativistic Electrons}}
  (\bibinfo{publisher}{American Institute of Physics}, \bibinfo{year}{1986}).

\bibitem[{\citenamefont{Niemi and Semenoff}(1986)}]{Niemi:1984vz}
\bibinfo{author}{\bibfnamefont{A.}~\bibnamefont{Niemi}} \bibnamefont{and}
  \bibinfo{author}{\bibfnamefont{G.}~\bibnamefont{Semenoff}},
  \bibinfo{journal}{Phys.Rept.} \textbf{\bibinfo{volume}{135}},
  \bibinfo{pages}{99} (\bibinfo{year}{1986}).

\bibitem[{\citenamefont{M.~Atiyah}(1973)}]{Atiyah1973}
\bibinfo{author}{\bibfnamefont{V.~P.} \bibnamefont{M.~Atiyah}},
  \bibinfo{journal}{I. Singer, Bull. London. Math. Soc.}
  \textbf{\bibinfo{volume}{5}}, \bibinfo{pages}{229} (\bibinfo{year}{1973}).

\bibitem[{\citenamefont{Son and Stephanov}(2008)}]{Son:2007ny}
\bibinfo{author}{\bibfnamefont{D.}~\bibnamefont{Son}} \bibnamefont{and}
  \bibinfo{author}{\bibfnamefont{M.}~\bibnamefont{Stephanov}},
  \bibinfo{journal}{Phys.Rev.} \textbf{\bibinfo{volume}{D77}},
  \bibinfo{pages}{014021} (\bibinfo{year}{2008}), \eprint{0710.1084}.

\bibitem[{\citenamefont{Inagaki et~al.}(2004)\citenamefont{Inagaki, Kimura, and
  Murata}}]{Inagaki:2003yi}
\bibinfo{author}{\bibfnamefont{T.}~\bibnamefont{Inagaki}},
  \bibinfo{author}{\bibfnamefont{D.}~\bibnamefont{Kimura}}, \bibnamefont{and}
  \bibinfo{author}{\bibfnamefont{T.}~\bibnamefont{Murata}},
  \bibinfo{journal}{Prog.Theor.Phys.} \textbf{\bibinfo{volume}{111}},
  \bibinfo{pages}{371} (\bibinfo{year}{2004}), \eprint{hep-ph/0312005}.

\bibitem[{\citenamefont{Mayaffre et~al.}(2014)\citenamefont{Mayaffre, Kramer,
  Horvatic, Berthier, Miyagawa, Kanoda, and Mitrovic}}]{Mayaffre:2014}
\bibinfo{author}{\bibfnamefont{H.}~\bibnamefont{Mayaffre}},
  \bibinfo{author}{\bibfnamefont{S.}~\bibnamefont{Kramer}},
  \bibinfo{author}{\bibfnamefont{M.}~\bibnamefont{Horvatic}},
  \bibinfo{author}{\bibfnamefont{C.}~\bibnamefont{Berthier}},
  \bibinfo{author}{\bibfnamefont{K.}~\bibnamefont{Miyagawa}},
  \bibinfo{author}{\bibfnamefont{K.}~\bibnamefont{Kanoda}}, \bibnamefont{and}
  \bibinfo{author}{\bibfnamefont{V.~F.} \bibnamefont{Mitrovic}},
  \bibinfo{journal}{Nat Phys} \textbf{\bibinfo{volume}{10}},
  \bibinfo{pages}{928} (\bibinfo{year}{2014}).

\bibitem[{\citenamefont{Thies}(2003)}]{Thies:2003zr}
\bibinfo{author}{\bibfnamefont{M.}~\bibnamefont{Thies}},
  \bibinfo{journal}{Phys.Rev.} \textbf{\bibinfo{volume}{D68}},
  \bibinfo{pages}{047703} (\bibinfo{year}{2003}), \eprint{hep-th/0303026}.

\bibitem[{\citenamefont{Yoshii et~al.}(2014)\citenamefont{Yoshii, Takada,
  Tsuchiya, Marmorini, Hayakawa et~al.}}]{Yoshii:2014fwa}
\bibinfo{author}{\bibfnamefont{R.}~\bibnamefont{Yoshii}},
  \bibinfo{author}{\bibfnamefont{S.}~\bibnamefont{Takada}},
  \bibinfo{author}{\bibfnamefont{S.}~\bibnamefont{Tsuchiya}},
  \bibinfo{author}{\bibfnamefont{G.}~\bibnamefont{Marmorini}},
  \bibinfo{author}{\bibfnamefont{H.}~\bibnamefont{Hayakawa}},
  \bibnamefont{et~al.} (\bibinfo{year}{2014}), \eprint{1404.3519}.

\end{thebibliography}
 
\end{document}